\documentclass[superscriptaddress,aps,preprintnumbers,showpacs,amssymb,prd,nofootinbib,preprint]{revtex4-1}

\usepackage{graphicx}
\usepackage{epsfig,amssymb,amsmath}

\pdfoutput=1
\usepackage{subfigure}
\usepackage[colorlinks=true,linkcolor=blue,citecolor=blue]{hyperref}
\usepackage{float}

\makeatletter
\def\Hline{%
\noalign{\ifnum0=`}\fi\hrule \@height 2pt \futurelet
\reserved@a\@xhline}
\makeatother

\newcommand{\beq}{\begin{equation}}
\newcommand{\eeq}{\end{equation}}
\newcommand{\bea}{\begin{eqnarray}}
\newcommand{\eea}{\end{eqnarray}}
\newcommand{\bear}{\begin{array}}
\newcommand {\eear}{\end{array}}
\newcommand{\bef}{\begin{figure}}
\newcommand {\eef}{\end{figure}}
\newcommand{\bec}{\begin{center}}
\newcommand {\eec}{\end{center}}

\def\GEV#1{10^{#1}{\rm\,GeV}}

\def\lrf#1#2{ \left(\frac{#1}{#2}\right)}
\def\lrfp#1#2#3{ \left(\frac{#1}{#2} \right)^{#3}}

\begin{document}
\draft
\tighten
\preprint{PNUTP-16/A12, TU-1025, IPMU16-0083, APCTP Pre2016-015, CTPU-16-16}
\title{
Topological Defects and nano-Hz Gravitational Waves \\
in Aligned Axion Models 
}
\author{
 Tetsutaro Higaki
}
\affiliation{Department of Physics, Keio University, Kanagawa 223-8522, Japan}
\author{
Kwang Sik Jeong
}
\affiliation{Department of Physics, Pusan National University, Busan 46241, Korea
}
\author{
Naoya Kitajima
}
\affiliation{Asia Pacific Center for Theoretical Physics, Pohang 790-784, Korea}
\author{
Toyokazu Sekiguchi
}
\affiliation{Center for Theoretical Physics of the Universe, Institute for Basic Science, Daejeon 34051, Korea}
\author{
Fuminobu Takahashi
}
\affiliation{Department of Physics, Tohoku University,
Sendai, Miyagi 980-8578, Japan}
\affiliation{Kavli IPMU (WPI), UTIAS,
The University of Tokyo,
Kashiwa, Chiba 277-8583, Japan}

%\date{\today}

%\vspace{2cm}

\begin{abstract}
We study the formation and evolution of topological defects in an aligned axion model with multiple Peccei-Quinn scalars, where the QCD 
axion is realized by a certain combination of the axions with decay constants much 
smaller than the conventional Peccei-Quinn breaking scale. 
When the underlying U(1) symmetries are spontaneously broken,
the aligned structure in the axion field space exhibits itself as a complicated string-wall network in the real space. 
We find that the string-wall network likely survives until the QCD phase transition if the number of the Peccei-Quinn scalars is greater than two. 
The string-wall system collapses during the QCD phase transition, producing a significant amount of gravitational waves in the nano-Hz range at present.
The typical decay constant is constrained to be below 
${\cal O}(100)$\,TeV  by the pulsar timing observations, and the constraint will be improved by a factor of $2$ in the future SKA observations.
\end{abstract}

\maketitle

%%%%%%%%%%%%%%%%%%%%%%%%%%%%%%%%%%%%%%
\section{Introduction}
\label{sec:intro}
%%%%%%%%%%%%%%%%%%%%%%%%%%%%%%%%%%%%%%

The Peccei-Quinn (PQ) mechanism is one of the most attractive solutions to the strong CP problem \cite{Peccei:1977ur,Peccei:1977hh}. In association with  spontaneous breakdown of a global U(1)$_{\rm PQ}$ symmetry, there arises a pseudo Nambu-Goldstone boson called the QCD axion \cite{Weinberg:1977ma,Wilczek:1977pj}. In the conventional scenarios, the axion decay constant $F_a$ is of order the U(1)$_{\rm PQ}$ breaking scale. The classical axion window is given by
\begin{equation}
\GEV{9} \lesssim F_a \lesssim \GEV{12},
\label{CAW}
\end{equation}
where the lower bound is due to the SN 1987A neutrino burst duration~\cite{Mayle:1987as,Raffelt:1987yt,Turner:1987by} and the upper bound is due to the cosmological abundance of the axion produced by the misalignment mechanism, barring fine-tuning of the initial misalignment~\cite{Preskill:1982cy,Abbott:1982af,Dine:1982ah}. The QCD axion is stable in a cosmological time scale, and is a plausible candidate for cold dark matter. See 
e.g. Refs.~\cite{Kim:1986ax,Kim:2008hd,Wantz:2009it,Ringwald:2012hr,Kawasaki:2013ae}  for reviews on the QCD axion and the related topics.

Axions are also known to appear in string theory in association with compactification of the extra dimension, and some of them may be so light that they play an important role in cosmology \cite{Conlon:2006tq,Svrcek:2006yi,Arvanitaki:2009fg,Cicoli:2012sz}.
In particular, a multi-axion inflation model attracted much attention, because
multiple axions with sub-Planckian decay constants can conspire to realize a light inflaton with the effective 
super-Planckian decay constant through the alignment mechanism~\cite{Kim:2004rp,Choi:2014rja,Higaki:2014pja,Higaki:2014mwa,Kappl:2014lra,Ben-Dayan:2014zsa,Long:2014dta}.
A peculiar structure of the charge assignment for the $N$ axions 
was noted in Ref.~\cite{Harigaya:2014rga}, and a concrete realization was given in Refs.~\cite{Choi:2015fiu,Kaplan:2015fuy}.
If there are many axions with various kinds of shift symmetry breaking terms, they may form an axion landscape~\cite{Higaki:2014pja,Higaki:2014mwa}, 
where  eternal inflation and the subsequent quantum tunneling or jump-up to one of the adjacent minima take place repeatedly. In the axion landscape, the slow-roll inflation may be realized via the alignment mechanism, and the inflaton potential is generally expected to have small modulations, which may lead to a sizable running 
of the spectral index~\cite{Kobayashi:2010pz,Czerny:2014wua}.
The vacuum structure of the axion landscape was studied in Refs.~\cite{Wang:2015rel,Masoumi:2016eqo}.

Recently, four of the present authors (TH, KSJ, NK, and FT) applied the alignment mechanism to the QCD axion, 
and studied its cosmological and phenomenological implications \cite{Higaki:2015jag,Higaki:2016yqk}. 
In this model, the QCD axion is given by a flat direction composed of multiple axions.
Interestingly, even if the decay constant of each axion is around the TeV scale,
the QCD axion decay constant can be enhanced up to the classical axion window (\ref{CAW}).
One of the axions or saxions in the model can explain the recently found diphoton excess in LHC~\cite{ATLAS,CMS:2015dxe}.\footnote{See Refs.~\cite{Chiang:2016eav,Gherghetta:2016fhp,DHHM} for
models of a visible axion as an explanation of the
diphoton excess.
}
In addition, the required high quality of the PQ symmetry \cite{Carpenter:2009zs,Fukuda:2015ana} can be naturally realized in this model \cite{Higaki:2016yqk}.

Cosmological evolution of the QCD axion can be quite involved even in the single axion case. Suppose that the U(1)$_{\rm PQ}$ symmetry is 
restored in the early Universe. This is the case if the inflation scale or the reheating temperature is sufficiently high. 
Then cosmic strings are formed when the U(1)$_{\rm {PQ}}$ is spontaneously broken.
Subsequently, the axion potential arises due to the QCD instanton effects, leading to the formation of domain walls  bounded 
by strings.
If the domain wall number of the QCD axion is unity, such a string-wall network soon collapses by emitting axions \cite{Hiramatsu:2012sc,Kawasaki:2014sqa}. 
On the other hand, if the domain wall number is greater than unity, the string-wall network is stable, which causes the cosmological 
disaster.\footnote{%%
There is an alternative way to form domain walls in a multi-axion scenario, where the system 
exhibits a chaotic behavior (axion roulette) during a mass level crossing between two axions \cite{Daido:2015bva,Daido:2015cba}.
}%% 

In the aligned axion model with $N$ axions, the associated U(1)$^N$ symmetries are likely restored in the early Universe,
because the symmetry breaking scale is much smaller than $F_a$.
The spontaneous breakdown of the U(1)$^N$ symmetries would result in the formation of a complicated  string-wall network, 
which reflects the aligned structure of the multiple axion fields.
The main purpose of this paper is to study the structure and evolution of the topological defects in the
aligned axion model.
As we shall see, in a case with two axions, an isolated string-wall system can be formed, which eventually collapses 
to a bundle of strings that can be regarded as the QCD axion string. In general, such a bundle of strings glued by walls
 exists as a solution corresponding to the QCD axion string, and in fact, their tensions are approximately same.
However, for  $N \geq 3$,  such a structure is rarely formed in the early Universe,
because the solution requires an exponentially large hierarchy in the number of strings of different types.
Instead, we expect that the string-wall network survives until a later time, and collapses during the QCD phase transition,
emitting gravitational waves. We derive an upper bound on the PQ breaking scale by the pulsar timing experiments~\cite{Lentati:2015qwp,Arzoumanian:2015liz,Verbiest:2016vem,Lasky:2015lej}.

The rest of this paper is organized as follows.
In Sec \ref{sec:aligned}, we  review the aligned QCD axion scenario. 
We discuss the formation and evolution of topological defects in the aligned QCD axion model in Sec \ref{sec:defect}.
Sec. \ref{sec:conc} is devoted to discussion and conclusions.

%%%%%%%%%%%%%%%%%%%%%%%%%%%%%%%%%%%%%%
\section{Aligned QCD axion}
\label{sec:aligned}
%%%%%%%%%%%%%%%%%%%%%%%%%%%%%%%%%%%%%%

To enhance the decay constant by the alignment mechanism, one needs multiple axions. 
Let us introduce $N$ axions, $\phi_1,\phi_2,\dots,\phi_N$, each of which
  respects the shift symmetry, 
\begin{equation}
    \phi_i \to \phi_i + C_i,
\end{equation}
where $C_i$ is a real transformation parameter. 
We assume that $N-1$ of the shift symmetries are explicitly broken down to their discrete subgroups, 
giving rise to the  potential for $(N-1)$ combinations of the axions. The remaining massless axion is to be identified 
with the QCD axion, and its anomalous coupling to gluons is generated by including extra PQ quarks
as in the Kim-Shifman-Vainshtein-Zakharov  axion models~\cite{Kim:1979if,Shifman:1979if} .
  The QCD axion is given by a certain combination of the $N$ axions, and it remains
massless until the QCD instanton effects are turned on.  In this set-up, the QCD axion decay constant
depends on how the $N-1$ shift symmetries are broken~\cite{Sikivie:1986gq}, and it
 can be significantly enhanced if the alignment takes place.

Here we adopt the clockwork axion model \cite{Choi:2015fiu,Kaplan:2015fuy} as a simple and concrete 
realization of the aligned axion model, and we apply it to the QCD axion following Ref.~\cite{Higaki:2015jag}.  The following discussion,
however,  can be straightforwardly applied to more general 
aligned axion models. In the clockwork axion model, the potential is given by
\beq \label{axionpot}
V = -\sum^{N-1}_{i=1} \Lambda_i^4 \cos \bigg( \frac{\phi_i}{f_i} + n_i \frac{\phi_{i+1}}{f_{i+1}} \bigg),
\eeq
for integer $n_i$, where $\Lambda_i$ and $f_i$ are a characteristic energy scale of the shift symmetry breaking and the decay constant of $\phi_i$, respectively. Since there are $N-1$ shift symmetry breaking terms, there remains a flat direction which is 
to be identified with the QCD axion. The QCD axion and its decay constant are given by
\begin{align} \label{eff_dec_const}
	a &\;\propto \;\sum^{N}_{i=1}(-1)^{i-1} \bigg(\prod^{N}_{j=i} n_j \bigg) f_i \phi_i, \\
	F_a &\;=\; \sqrt{\sum_{i=1}^{N} f_i^2 \left( \prod_{j=i}^{N} n_j^2\right)},
\end{align}
where we have defined $n_N = 1$ for notational convenience.
Suppose that all the decay constants $f_i$ and the integers $n_i$ are comparable to each other, i.e. $f_i \simeq f$ and $n_i \simeq n$. Then,
the largest contribution to the QCD axion comes from $\phi_1$,  and  the QCD axion decay constant is exponentially 
 enhanced, $F_a \sim e^{N \ln n} f$, for large $N$.
For instance, if $f_i \simeq f = {\cal O}(1)$\,TeV, one needs an  enhancement of order $10^{6-9}$  for $F_a$ to be in the classical axion
window (\ref{CAW}), which can be realized for e.g. $n=3$ and $N = 13 - 19$.

One of the virtues of the aligned QCD axion model is that the high quality of the PQ symmetry is naturally explained~\cite{Higaki:2015jag,Higaki:2016yqk}.
This is because the actual symmetry breaking scale is much smaller than $F_a$, and so, any Planck-suppressed PQ breaking terms give only
suppressed contributions to the QCD axion potential. In some case, such Planck-suppressed PQ breaking terms have interesting implications
for the QCD axion dynamics~\cite{Higaki:2016yqk}.\footnote{See also Ref.~\cite{Jaeckel:2016qjp} for the study of  the pseudo Nambu-Goldstone dark matter
in a potential with small modulations.}

Another cosmological implication is that the PQ symmetry is easily restored in the early Universe.  To see this, 
let us now consider a possible UV completion, where each axion $\phi_i$ is originally embedded in a phase component of a complex scalar field, 
$\Phi_i$. The model is based on global U(1)$^N$ symmetries and the scalar potential has a typical form of
\beq
\label{potential}
	V = \sum^N_{i=1} \bigg( -m_i^2 |\Phi_i|^2 + \lambda_i |\Phi_i|^4 \bigg).
\eeq
In addition, the scalars generically acquire a Hubble-induced mass or thermal mass in the early Universe.
Then, if the Hubble parameter or the temperature is higher than $m_i$, the scalars are stabilized at the origin, and the U(1)$^N$ symmetries are restored. 
As the Universe expands, $\Phi_i$ becomes tachyonic at the origin, and develops 
a vacuum expectation value, $|\Phi_i| = f_i/\sqrt{2}$ with $f_i = m_i/\sqrt{\lambda_i}$. Then, 
$N$ massless axions $\phi_{1,\cdots, N}$ appear as Nambu-Goldstone bosons. 
At the same time,  cosmic strings are produced in association with a non-trivial topological configuration of $\phi_i$.
Later on, the $(N-1)$ symmetry breaking terms become important and the $(N-1)$ axions acquire masses.
As an example, we may adopt the following  renormalizable potential \cite{Kaplan:2015fuy},
\beq \label{clockwork_comp}
	\Delta V = \sum^{N-1}_{i=1} \epsilon_i \Phi_i \Phi^3_{i+1} + {\rm h.c.},
\eeq
where $\epsilon_i$ is an order parameter;  $\epsilon_i \ll 1$ implies that the corresponding U(1) symmetry is a relatively
good symmetry.
This potential generates the axion potential (\ref{axionpot}) with $n_i = 3$.
As a result,  domain walls appear stretching between the cosmic strings. The strings and walls form a complicated
string-wall network.
We will study the nature of these topological defects in the next section.

%%%%%%%%%%%%%%%%%%%%%%%%%%%%%%%%%%%%%%
\section{Topological defects and gravitational waves}
\label{sec:defect}
%%%%%%%%%%%%%%%%%%%%%%%%%%%%%%%%%%%%%%

Here we study the formation and evolution of cosmic strings and domain walls in the aligned axion model.
First we discuss the structure of string-wall network and its correspondence with the ordinary QCD axion string.
Then we investigate their evolution based on numerical simulations.
Lastly, we evaluate the gravitational waves emitted in the process of the domain wall collisions during the
QCD phase transition.

\subsection{Cosmic strings and domain walls}
\label{subsec:string-wall}
In the aligned QCD axion model,  the actual symmetry breaking scales, $f_i$, are much smaller than $F_a$, and so,
the U(1)$^N$ symmetries are easily restored in the early Universe.
When the  U(1)$^N$ symmetries are spontaneously broken, the radial component of each complex scalar field develops a nonzero vacuum expectation 
value, while the phase component is randomly distributed in space.  There appear
 $N$ kinds of cosmic strings corresponding to a non-trivial topological configuration of $\phi_{1,\cdots,N}$.
Consider a string of $\Phi_i$ along the $z$-axis. Then,
one can express the complex scalar field around the string in a cylindrical coordinate system as
\beq \label{eq:asympt}
\Phi_i = \frac{f_i}{\sqrt{2}} e^{i \phi_i/f_i} = \frac{f_i}{\sqrt{2}} e^{i w_i \theta},
\eeq
where $\theta$ is the angular coordinate, $w_i$ is the winding number and we have neglected the radial component of $\Phi_i$.
Most of cosmic strings realized in the spontaneous symmetry breaking 
 have $w_i = \pm 1$, and so, we will focus on this case, and
denote the (anti-)string with $w_i=1(-1)$ by $S_i$ ($\bar{S}_i$) in the following. 
The energy per unit length, i.e., the sting tension,  for an infinitely long global string is given by the sum of the potential energy stored inside the string core
and the gradient energy outside the string. The former is roughly $\mu_{{\rm core},i} \sim f_i^2$, and the latter is logarithmically 
divergent and gives the dominant contribution.
 For each cosmic string, the tension  is estimated as
\beq \label{eq:tension}
	\mu_i \sim \mu_{\rm core} + \int^R_\delta \bigg|\frac{1}{r} \frac{\partial \Phi_i}{\partial \theta_i} \bigg|^2 2\pi r dr \approx \pi w_i^2 f_i^2 \ln \bigg(\frac{R}{\delta} \bigg),
\eeq
where $\delta \sim m_i^{-1}$ is the typical core radius, and
$R$ is the cutoff length corresponding to the distance between strings, which is usually the Hubble radius.
Those $N$ kinds of strings individually follow the scaling law until the domain walls are formed. 

Later on, the shift symmetry breaking terms such as (\ref{clockwork_comp}) become important, 
and the $N-1$ U(1) symmetries are explicitly broken down to their discrete subgroups, generating
discrete minima for the $(N-1)$ axions.   As a result, domain walls appear between strings. 
There are two kinds of domain walls, $W_{(i,i+1)}$ and 
$W_{(i, i)}$, where the former stretches between ${\bar S}_i$ and $S_{i+1}$ with $i = 1, \cdots, N-1$, and
the latter between the anti-string $\bar{S}_i$  and string $S_i$.
For the potential (\ref{axionpot}), a single wall  is attached to $S_1$,
$n_{i-1} + 1$ walls are attached  to $S_i$ with $i = 2, \cdots, N-2$, and $n_{N-1}$ walls are attached to $S_N$. 
The composition of the walls depends on
the initial string configuration, and it also evolves with time as strings and walls annihilate.
Specifically, 
$S_i$ with $i = 2, \cdots, N-2$ has $n_{i-1} + 1$ walls whose composition is expressed by
\beq
a_i W_{(i,i+1)} +  b_{i-1} W_{(i,i-1)}+(n_{i-1}+1-a_i-b_{i-1}) W_{(i,i)}
\label{comp}
\eeq
with $a_i = 0, 1$ and $b_i = 0, \cdots,n_{i}$.
Those strings and walls form a complicated string-wall system, whose evolution is 
numerically studied  later in this section.

\subsection{Cosmic string bundles as QCD axion strings}
\label{subsec:bundle}

Here we show that there is a special configuration in which many strings $S_{1,\cdots,N}$ glued by domain walls form
an isolated string bundle. The effective tension of the bundle is of the same order of the QCD axion string in a usual
axion model without the alignment.

\begin{figure}[tp]
\centering
\subfigure[~$N = 2$] {
\includegraphics [width = 5.0cm, clip]{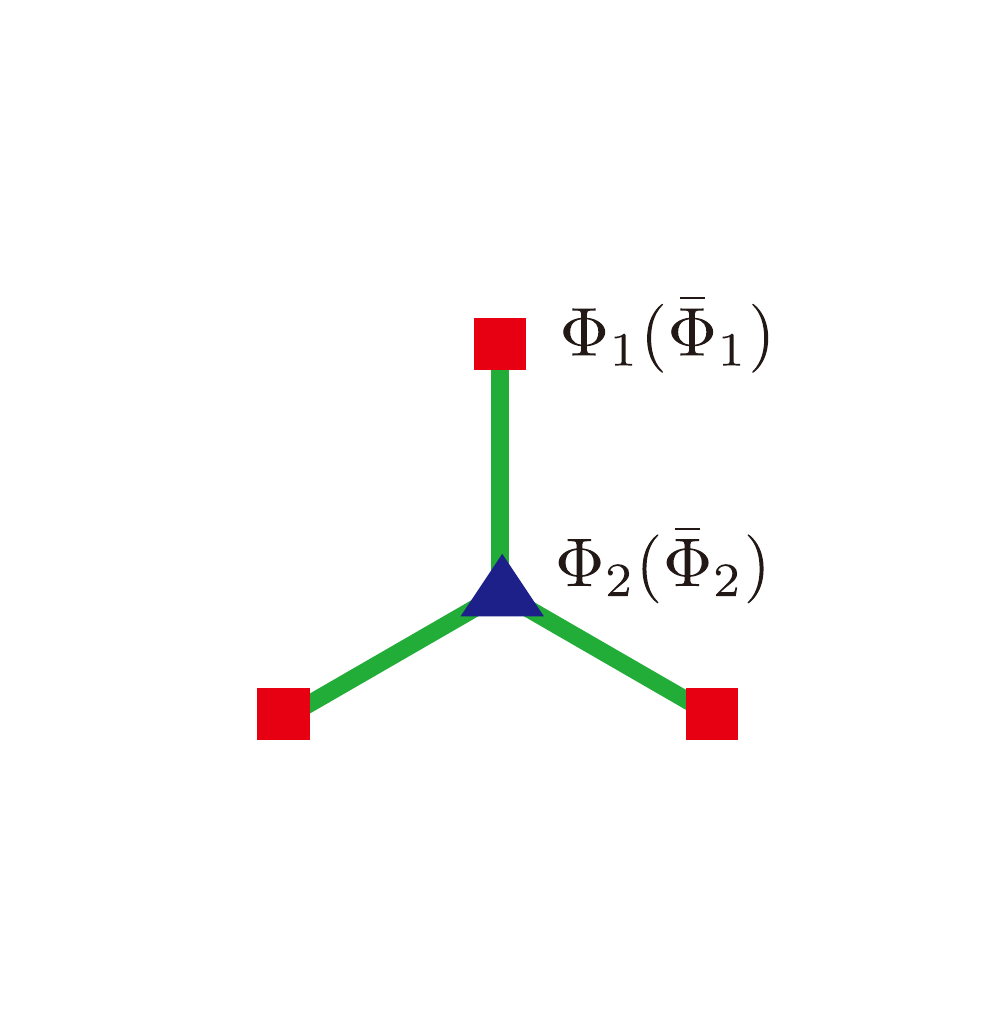}
\label{subfig:N2}
}
\subfigure[~$N = 3$]{
\includegraphics [width = 5.0cm, clip]{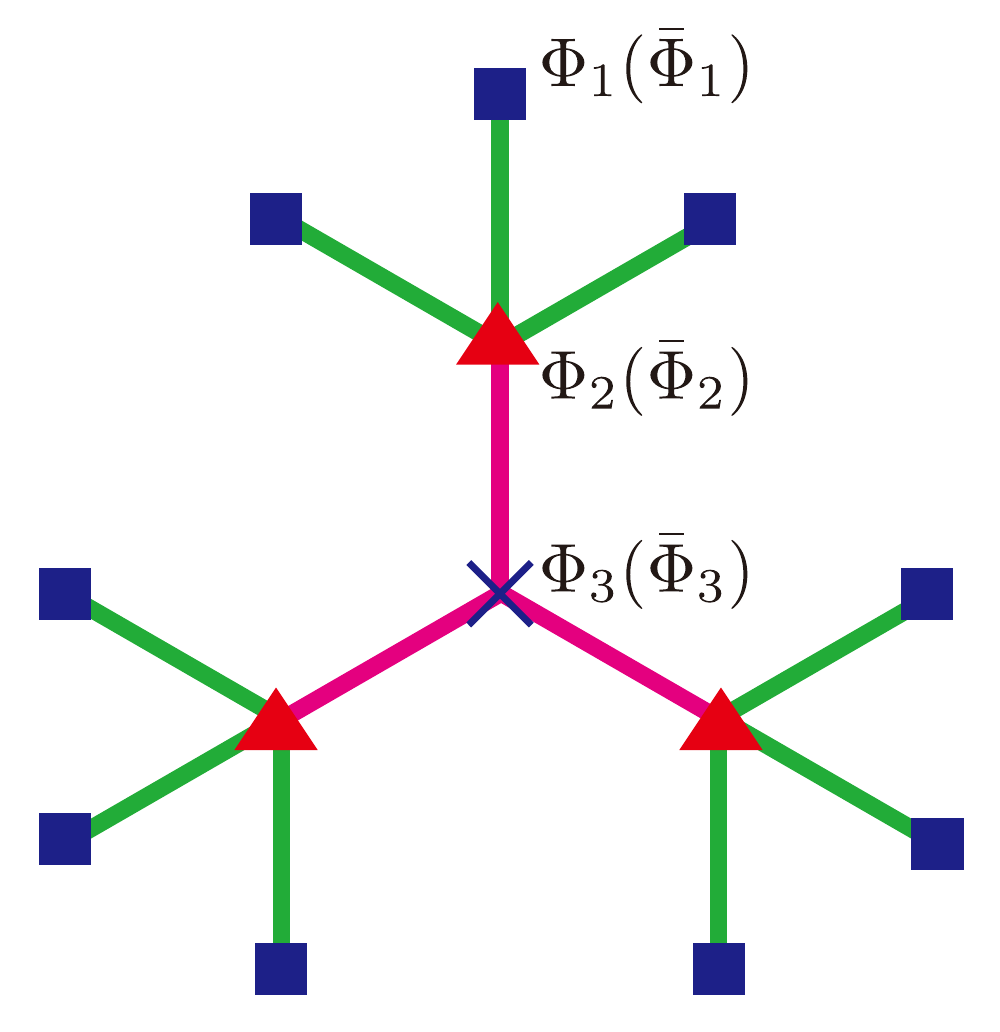}
\label{subfig:N3}
}
\caption{
Isolated string-wall structure in the clockwork axion model with $N=2$ (left) and $N=3$ (right), where we set $n_i=3$.
The squares, triangles and cross-marks represent cosmic 
strings for $\Phi_1$, $\Phi_2$ and $\Phi_3$ respectively and green and magenta lines represent domain walls connecting each string.
}
\label{fig:string_struct}
\end{figure}

Let us begin with the case of $N=2$ and adopt the renormalizable potential (\ref{clockwork_comp}). 
In this case, one possible configuration is such that three $\bar{S}_1$ strings are connected to an $S_2$ string by the walls $W_{12}$,
as schematically illustrated in Fig.~\ref{subfig:N2}.
Once such a configuration is formed, each string gets attracted to each other by the domain wall tension, leading to 
a single string bundle composed of $S_2+3 \bar{S}_1$ glued by the walls $W_{12}$. Taking account of the fact that the effective winding number  
$w_1$  is equal to $-3$, the effective tension of such a string bundle is 
\beq
	\mu_{\rm eff} \simeq \pi( 3^2 f_1^2 + f_2^2) \ln \bigg( \frac{R}{\delta} \bigg) = \pi F_a^2 \ln \bigg( \frac{R}{\delta} \bigg),
\eeq
where we have used the effective decay constant, $F_a = \sqrt{ 3^2 f_1^2 + f_2^2}$ (see the relation (\ref{eff_dec_const})), in the second equality,
and we have neglected subdominant contributions from the string cores and the walls.
The tension is consistent with that of the QCD axion string with the winding number equal to unity.
The above argument can be straightforwardly applied to the case of $N \geq 3$. An isolated string bundle is composed of
$S_N + 3 \bar{S}_{N-1}+3^2 S_{N-2}+ \cdots + 3^{N-1} S_1 (\bar{S}_1)$ for odd (even) $N$.
See Fig.~\ref{subfig:N3} in the case of $N=3$. The effective tension is in general
\beq
	\mu_{\rm eff} \simeq \pi( 3^{2(N-1)} f_1^2 + \cdots + 3^2 f_{N-1}^2+f_N^2) \ln \bigg( \frac{R}{\delta} \bigg) = \pi F_a^2 \ln \bigg( \frac{R}{\delta} \bigg).
\eeq
Thus, the string bundle can be regarded as the usual QCD axion string with the decay constant $F_a$.

Let us note here that the aligned structure in the axion field space appears in the real space as the string bundle.
The string bundle contains exponentially many strings, and as a result, the effective winding number of each string
becomes exponentially large. This enhances the string tension by a factor of $F_a^2/f^2$.

%%%%%%%%%%%%%%% MULTI-FIGURE  %%%%%%%%%%%%%%%
%%%%%%%%%%%%%%%%%%%%%%%%%%%%%%%%%%%%%%%

To reinforce the estimation above, here we numerically analyze the structure of an isolated string bundle for $N=2$.
For this purpose, we have computed a static solution of $\Phi_i$ that minimizes the action with 
a potential consisting of Eqs. \eqref{eff_dec_const} and \eqref{clockwork_comp}.
The action is computed on a 2-dim lattice with the number of grids $N_{\rm grid}=512^2$. 
The box size is taken to be about 70 times larger than the typical length scale of strings $({\sqrt\lambda_i} f_i)^{-1}$.
To obtain an isolated string bundle, we have adopted a Dirichlet boundary condition, 
so that along the boundary Eq.~\eqref{eq:asympt} is satisfied.
The minimization of nonlinear actions with such a large $N_{\rm grid}$ is not trivial.
In this paper we have adopted the {\it scheduled relaxation Jacobi} method 
\cite{Yang:2014,Adsuara:2015eds}.

\begin{figure}[tp]
\centering
\includegraphics [width = 18.0cm, clip]{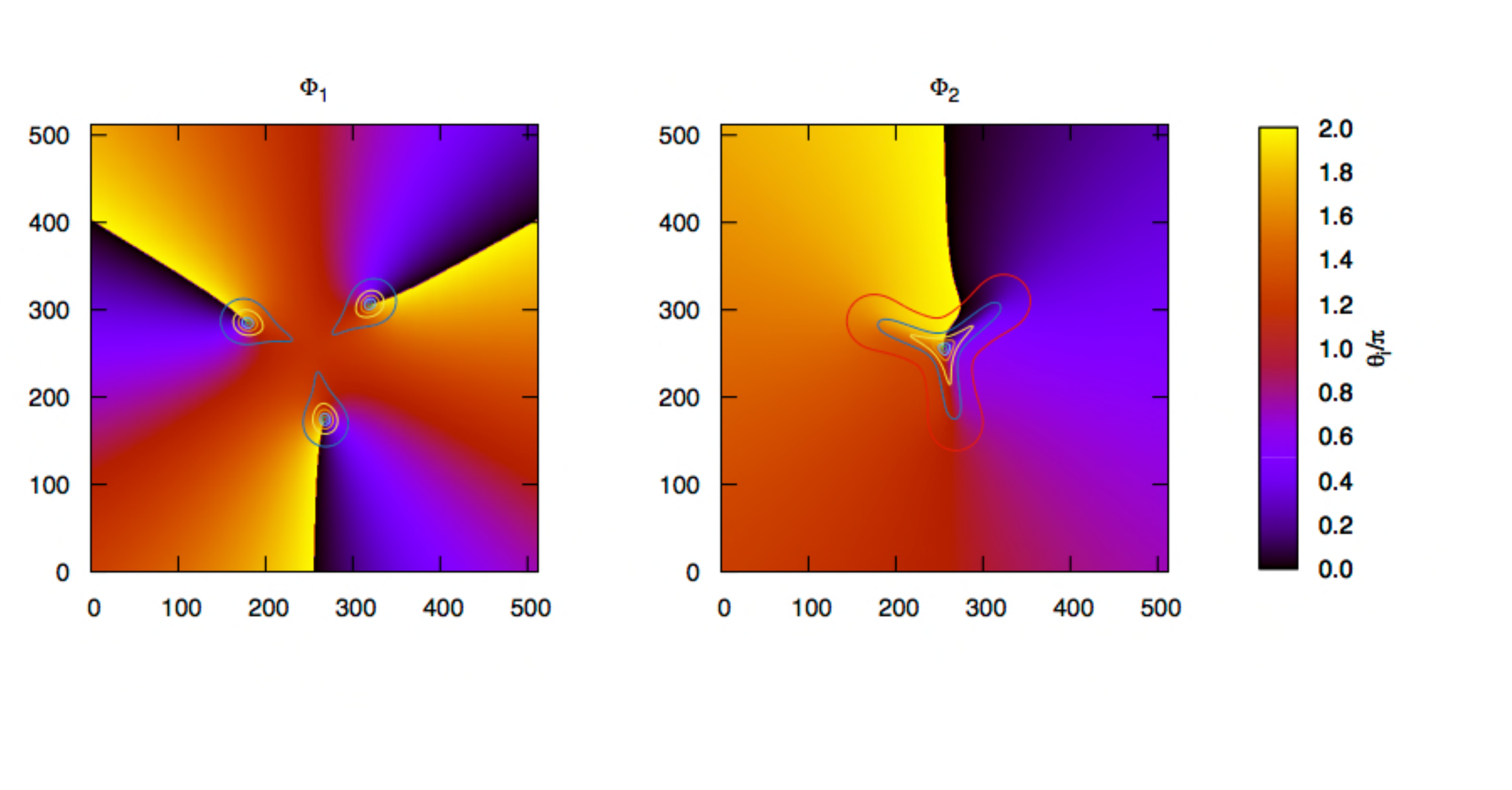}
\caption{
Field configuration around an isolated string bundle for $N=2$ with $f_1=f_2=f$ and $\epsilon=0.25$.
Left and right panels respectively show $\Phi_1$ and $\Phi_2$.
Contours indicate $|\Phi_i|/(f/\sqrt2)$, whose values are incremented from 0 (innermost) by 0.2, while
color scales indicate $\theta_i/\pi\in[0,2)$.
}
\label{fig:string_config}
\end{figure}

Fig.~\ref{fig:string_config} shows the configurations of $\Phi_1$ and $\Phi_2$.
Here we have assumed $f_1=f_2=f$ and $\epsilon=0.25$. 
First of all, it can be clearly seen that there are three distinct $S_1$ strings, 
which surround the $\bar S_2$ string at the center. This is consistent with 
the schematic picture in Fig.~\ref{subfig:N2}.
We also note that the contours indicating $|\Phi_2|$ are elongated towards the cores of the $S_1$ strings.
This is because domain walls stretch between $S_1$ and $\bar S_2$ strings. For the potential energy in 
$\Delta V$ of Eq.~\eqref{clockwork_comp} to be reduced, $|\Phi_2|$ is required to be nonzero along the domain walls.
We note that in the right panel in Fig.~\ref{fig:string_config}, the discontinuity in the color scale corresponding to $\theta_2=0$ 
appears wiggling, which results from minimizing the potential energy around domain walls.
While it is not as apparent as $\theta_2=0$,  loci corresponding to $\theta_2=2/3\pi$ and $\theta_2=4/3\pi$ 
are also wiggling in a similar way.

\begin{figure}[tp]
\centering
\includegraphics [width = 10.0cm, clip]{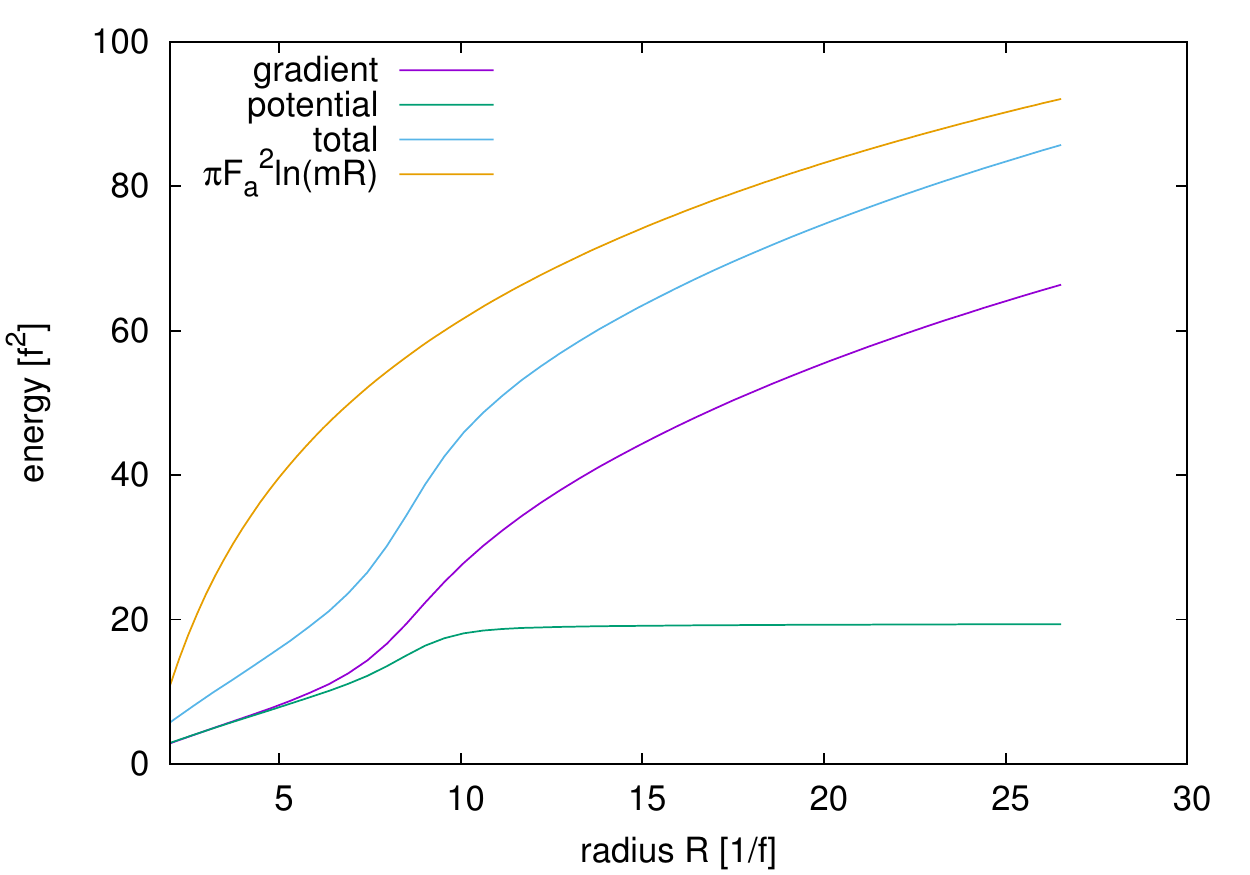}
\caption{
Energy stored inside the radius from the center of the string bundle in Fig.~\ref{fig:string_config}.
The green and purple lines are respectively the gradient and potential energy, while 
the cyan one is the sum of them. 
For reference, the yellow line is depicted to show 
Eq.~\eqref{eq:tension} with $\delta$ set to $1/m=\sqrt2/f$.
}
\label{fig:string_tension}
\end{figure}

From our numerical calculation, we can also compute the tension of the string bundle.
In Fig.~\ref{fig:string_tension}, we plot the energy stored inside the radius
from the center of the bundle. Because the gradients of the phases $\theta_i$ dominate at large radii,
the energy approximately logarithmically increases and its asymptotic 
behavior converges into Eq.~\eqref{eq:tension}.

%%%%%%%%%%%%%%% MULTI-FIGURE  %%%%%%%%%%%%%%%
%%%%%%%%%%%%%%%%%%%%%%%%%%%%%%%%%%%%%%%

%%%%%%%%%%%%%%% MULTI-FIGURE  %%%%%%%%%%%%%%%
%%%%%%%%%%%%%%%%%%%%%%%%%%%%%%%%%%%%%%%

%%%%%%%%

%%%%%%%%%%%%%%%%%%%%%%%%%%%%%%%%%%%%%%%%%%%%%%%%%
\subsection{Formation and evolution of string-wall network}
\label{subsec:probability}

Now the question is if isolated string bundles are produced after the domain walls appear in the early Universe.
As we shall argue below, such isolated string bundles are unlikely to be formed if $N \geq 3$.

Starting from the random initial condition,
cosmic strings are necessarily produced  after the spontaneous breakdown of the $N$ U(1) symmetries. 
The cosmic strings are considered to follow the scaling law in a few Hubble time, and the number of each
string per the Hubble horizon is of order unity. 
At a later time, the shift symmetry breaking terms generate discrete minima for the $(N-1)$ axions, and
domain walls are formed  between two strings. 

Let us take one $S_N$ string, and consider a whole object
connected to it by various kinds of walls and strings. 
Assume that such an object contains a finite number of strings.
As we start from the $S_N$ string, each string contained in this object is identified with either $S_i$ or $\bar{S}_i$ 
with $i = 1, \cdots, N$. Namely, the orientation of the strings is determined by how they are connected
to the $S_N$ string.  The number of strings is monotonically decreasing with time unless pairs of string and anti-string
are created.
If the difference between the number of $S_N$ strings
and that of ${\bar S}_N$ strings contained in this object is equal to unity, i.e., if
\beq
\label{SN}
\left|\#(S_N) - \#(\bar{S}_N)\right| =  1,
\eeq
such an object will collapse into an isolated string bundle at a sufficiently late time when the whole object
is contained in a Hubble horizon. Since there is no essential difference between $S_N$ and ${\bar S}_N$
and their number density is comparable, 
the above condition (\ref{SN}) is easily satisfied if such an object 
with a finite number of strings is formed.
However, the object must have a larger bias in the number of the strings of the other types. 
This can be understood by noting that the object must contain much more $S_1(\bar{S}_1)$ strings
than  $\bar{S}_N(S_N)$ strings for odd (even) $N$, as it collapses into 
$S_N + 3 \bar{S}_{N-1}+3^2 S_{N-2}+ \cdots + 3^{N-1} S_1 (\bar{S}_1)$ at the end of the day.
That is to say, the number of strings in the object must satisfy
\beq
\left|\#(S_i) - \#(\bar{S}_i)\right| =  3^{N-i}
\eeq
for $1 \leq i \leq N$. However, starting with the random initial condition, 
the number of strings and anti-strings in a part of the whole string-wall network
typically satisfy
\beq
\left|\#(S_i) - \#(\bar{S}_i)\right| = {\cal O}(1).
\eeq
Therefore it is extremely unlikely that the required huge bias in the number of strings is realized. 
We thus conclude that such an object contains most likely 
an infinite number of strings
for large $N$, and no isolated
objects are produced.\footnote{
This may imply that, even if the scalar potential for multiple PQ scalars is complicated,
such isolated structure will be likely formed unless the alignment is realized. 
} Indeed, as we shall see below, we could not find any isolated structure in the
numerical simulations with the $N=3$ case (see Fig.~\ref{lattice3D_n3}).
On the other hand, in the case of $N=2$, the required bias is not huge, and isolated string bundles
are formed  (see Fig.~\ref{lattice3D_n2}).

Even if isolated string bundles are not formed,
the string-wall network will never disappear until the QCD instanton effects are turned on. This can be understood as follows.  Note that
only $N-1$ of the original $N$ U(1) symmetries are explicitly broken, and there remains a massless degree of freedom
corresponding to the QCD axion, $a$.  Before the domain wall formation,
all possible field values of  the $N$ axions are realized in space, and there is no bias in their distribution. However,
if the string-wall network disappeared, it would imply that the distribution of the QCD axion
is topologically trivial everywhere. This is the case if only a part of its field space is realized, which is inconsistent
with the initial condition.  Indeed,  the QCD axion has
a non-trivial topological configuration around the $S_N$ string, which would be the core of isolated
string bundle. In other words, the $S_N$ strings cannot completely disappear in the Universe until the QCD phase
transition, irrespective of whether isolated string bundles with $S_N$ being the core are formed or 
a complicated string-wall network remains around it.

In summary, it is highly unlikely that isolated string bundles are formed in the Universe for large $N$,
as long as the initial condition is random.
The string-wall network in the aligned axion model likely remains until the QCD phase transition.

%%%%%%%%%%%%%%%%%%%%%%%%%%%%%%%%%%%%%%
\subsection{Numerical simulation of the string-wall network}
\label{subsec:numerical}
%%%%%%%%%%%%%%%%%%%%%%%%%%%%%%%%%%%%%%

\begin{figure}[tp]
\centering
\subfigure[]{
\includegraphics [width = 5.0cm, clip]{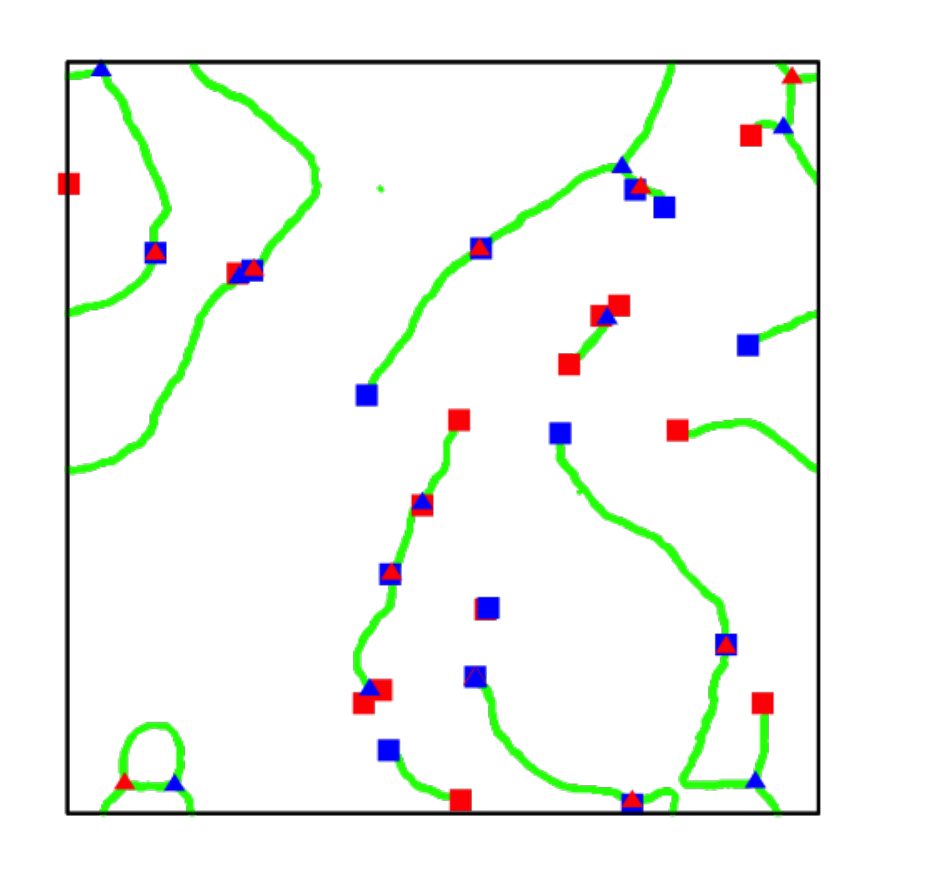}
\label{lattice_n2_1}
}
\subfigure[]{
\includegraphics [width = 5.0cm, clip]{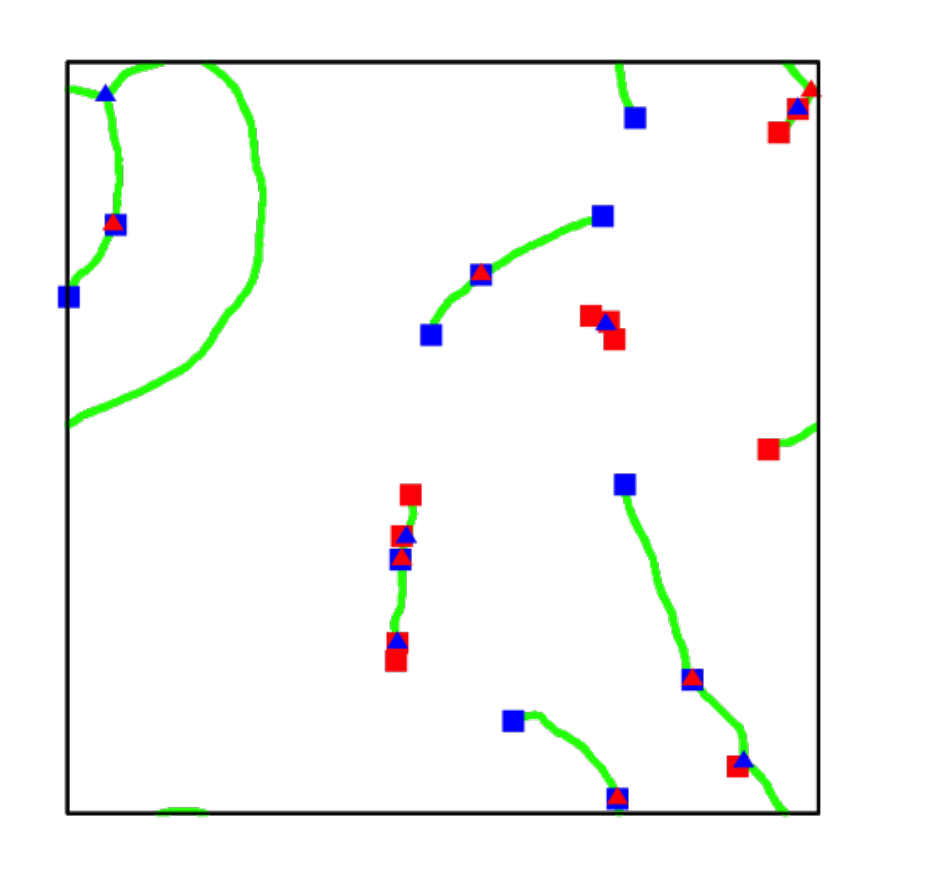}
\label{lattice_n2_2}
}
\subfigure[]{
\includegraphics [width = 5.0cm, clip]{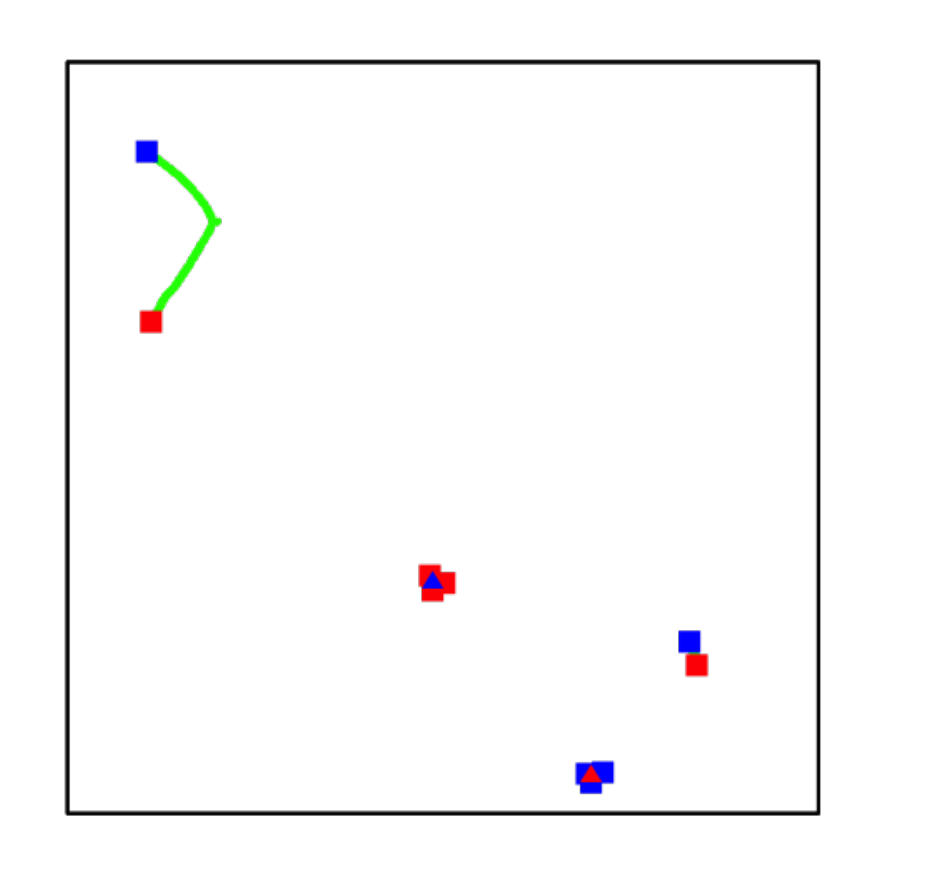}
\label{lattice_n2_3}
}
\caption{
Two-dimensional lattice simulation for $N=2$. Time evolves from left to right.
The red (blue) square and triangle points represent 
$S_1$ ($\bar{S}_1$) and $S_2$ ($\bar{S}_2$) respectively and the green line represents 
$W_{12}$.
}
\label{lattice_n2}
\end{figure}

\begin{figure}[tp]
\centering
\subfigure[]{
\includegraphics [width = 5.0cm, clip]{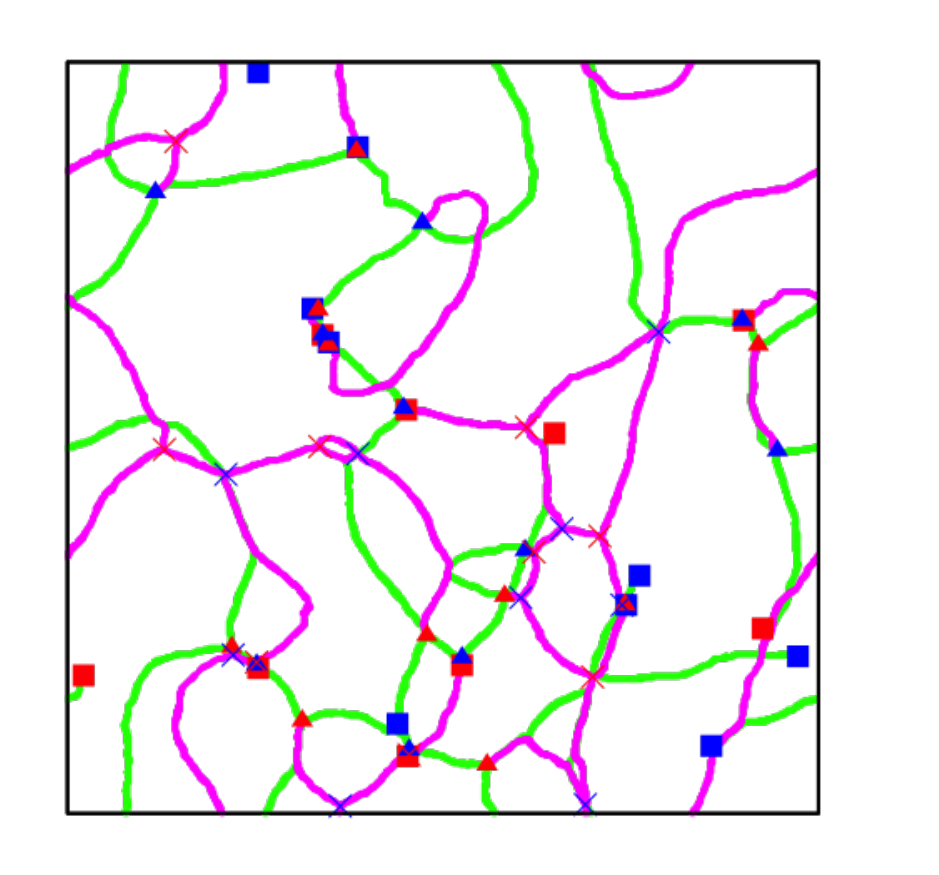}
\label{lattice_n3_1}
}
\subfigure[]{
\includegraphics [width = 5.0cm, clip]{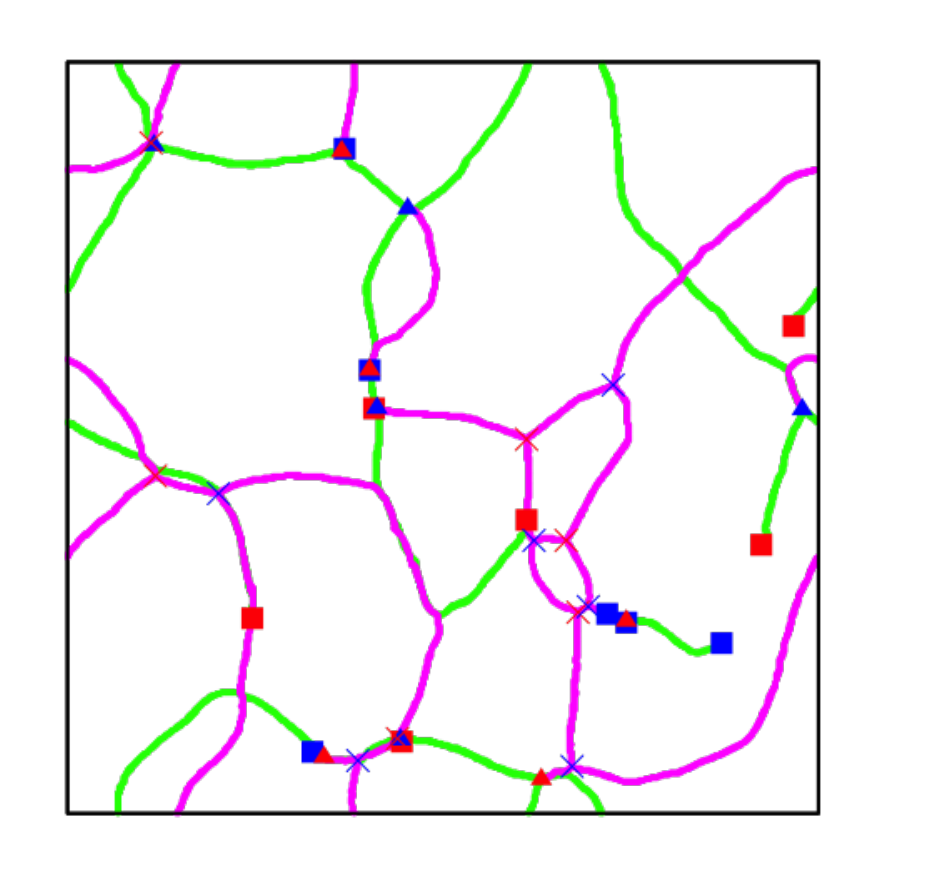}
\label{lattice_n3_2}
}
\subfigure[]{
\includegraphics [width = 5.0cm, clip]{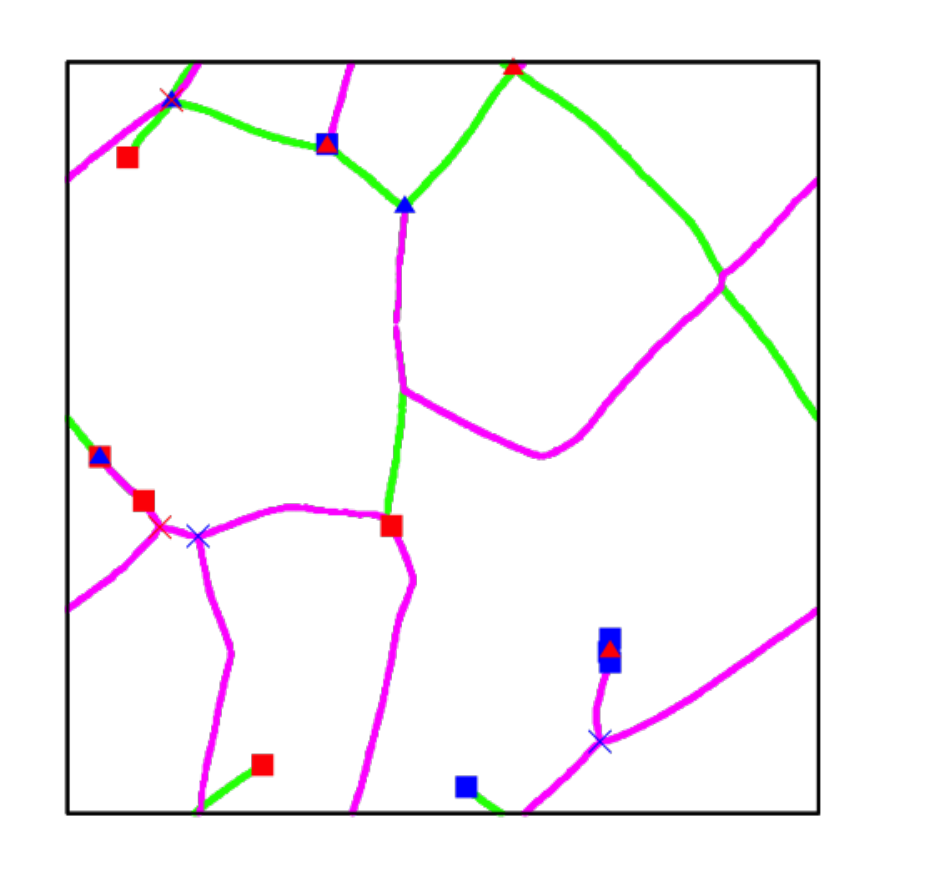}
\label{lattice_n3_3}
}
\caption{
Same as Fig.~\ref{lattice_n2} but for $N=3$.
The red (blue) x-mark represents $S_3$ ($\bar{S}_3$) and the magenta line represents $W_{23}$.
}
\label{lattice_n3}
\end{figure}

\begin{figure}[tp]
\centering
\subfigure[]{
\includegraphics [width = 5.0cm, clip]{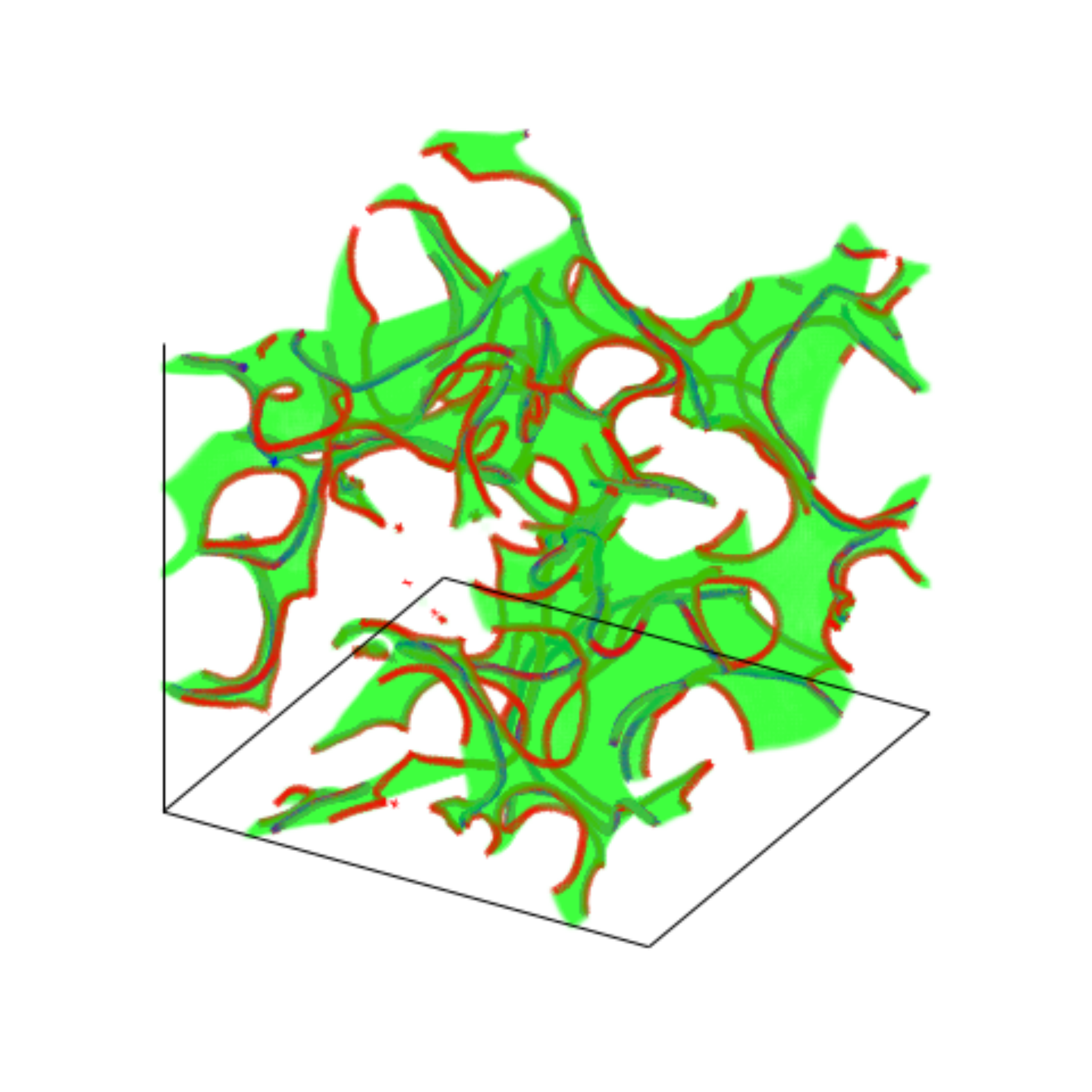}
\label{lattice3D_n2_1}
}
\subfigure[]{
\includegraphics [width = 5.0cm, clip]{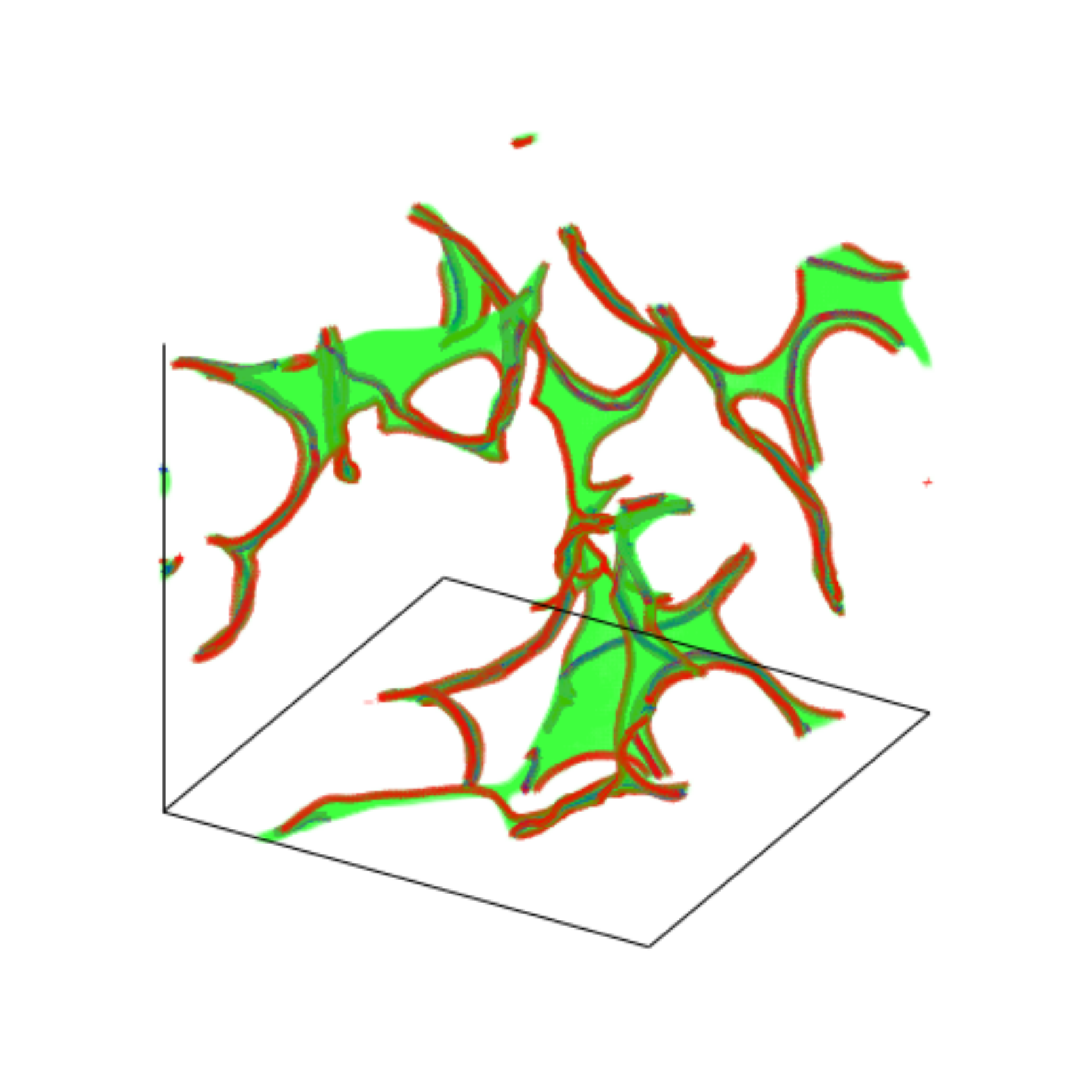}
\label{lattice3D_n2_2}
}
\subfigure[]{
\includegraphics [width = 5.0cm, clip]{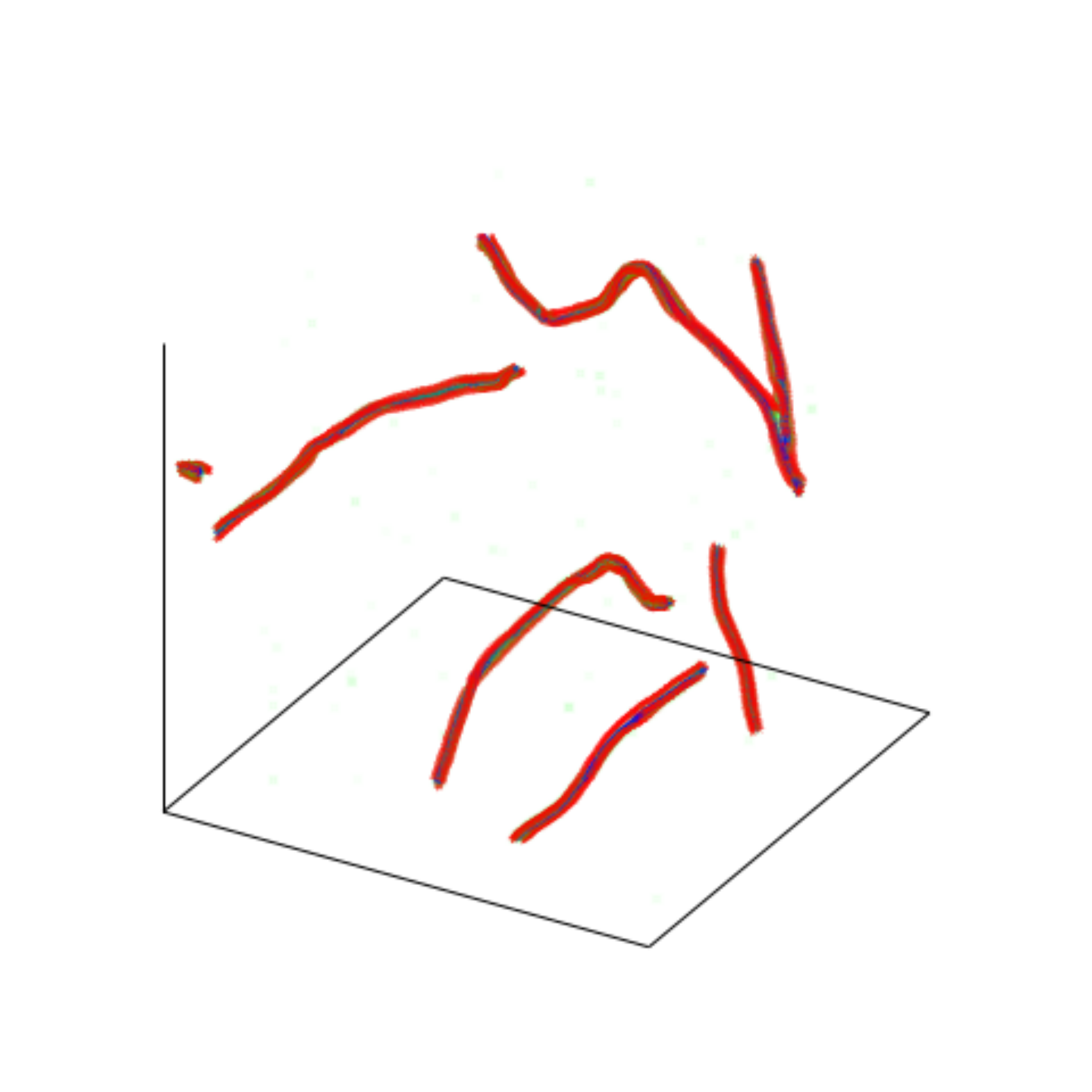}
\label{lattice3D_n2_3}
}
\caption{
Three-dimensional lattice simulations for $N=2$. Time evolves from left to right.
The red and blue lines represent $S_1$ and $S_2$ respectively and green region represents $W_{12}$.
}
\label{lattice3D_n2}
\end{figure}

\begin{figure}[tp]
\centering
\subfigure[]{
\includegraphics [width = 5.0cm, clip]{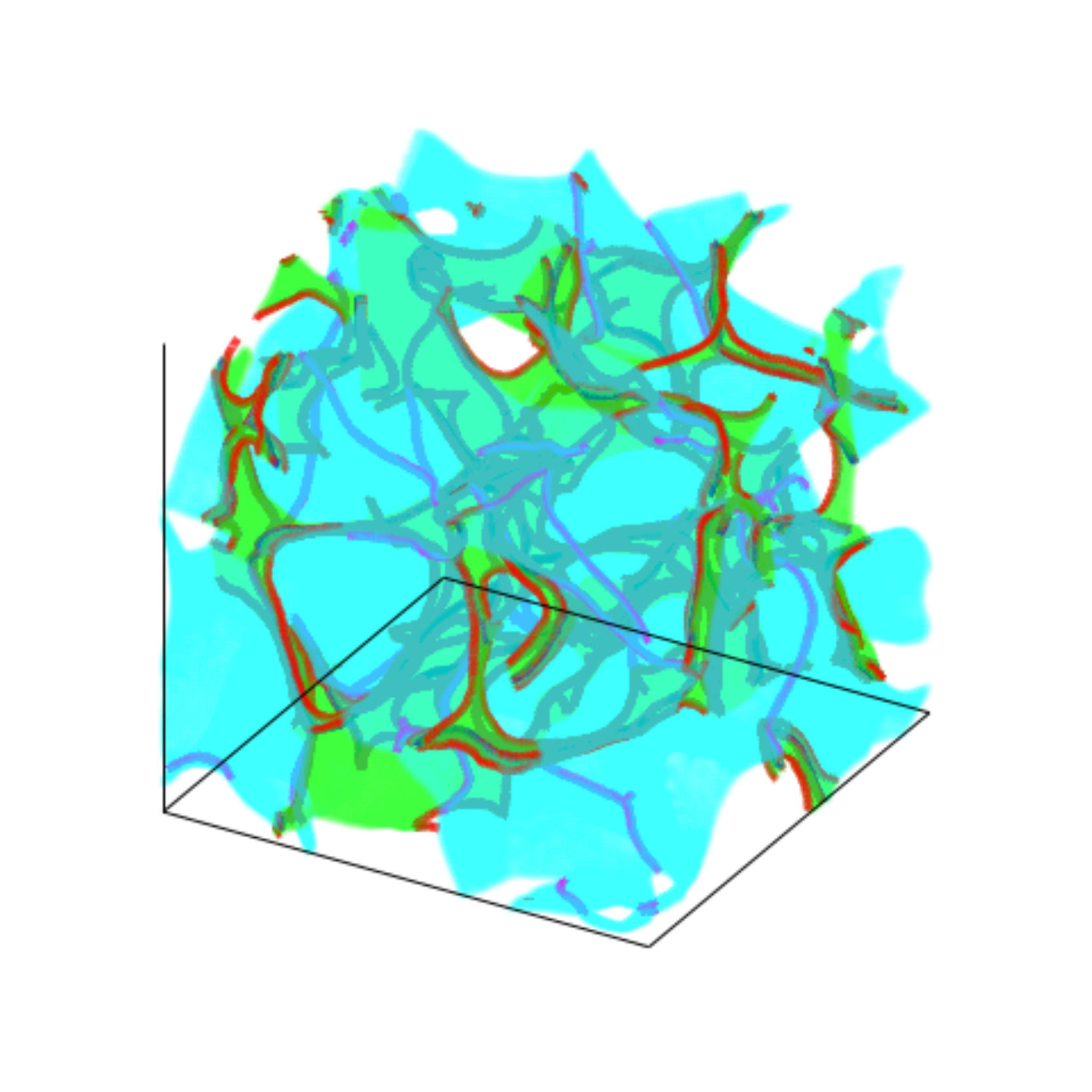}
\label{lattice3D_n3_1}
}
\subfigure[]{
\includegraphics [width = 5.0cm, clip]{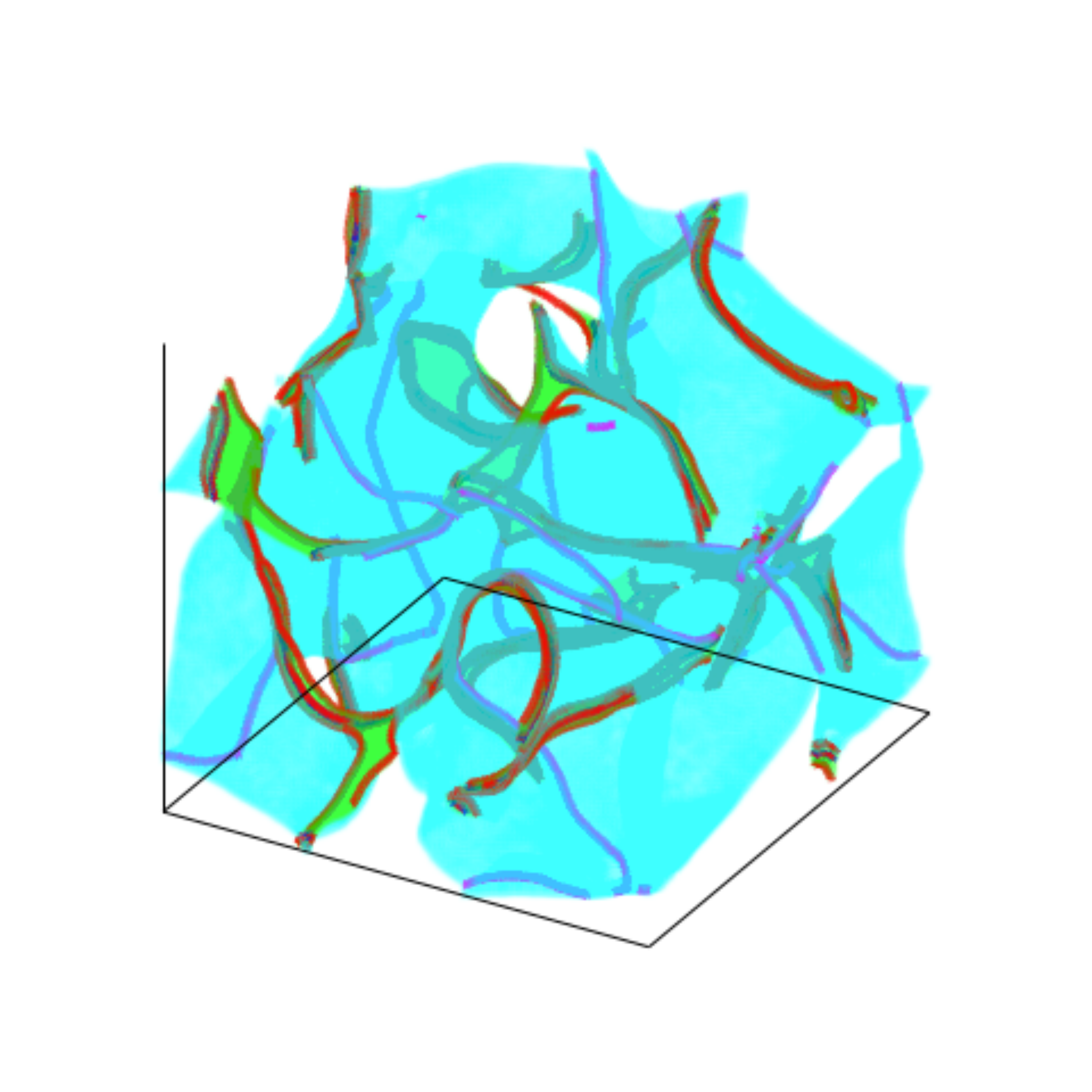}
\label{lattice3D_n3_2}
}
\subfigure[]{
\includegraphics [width = 5.0cm, clip]{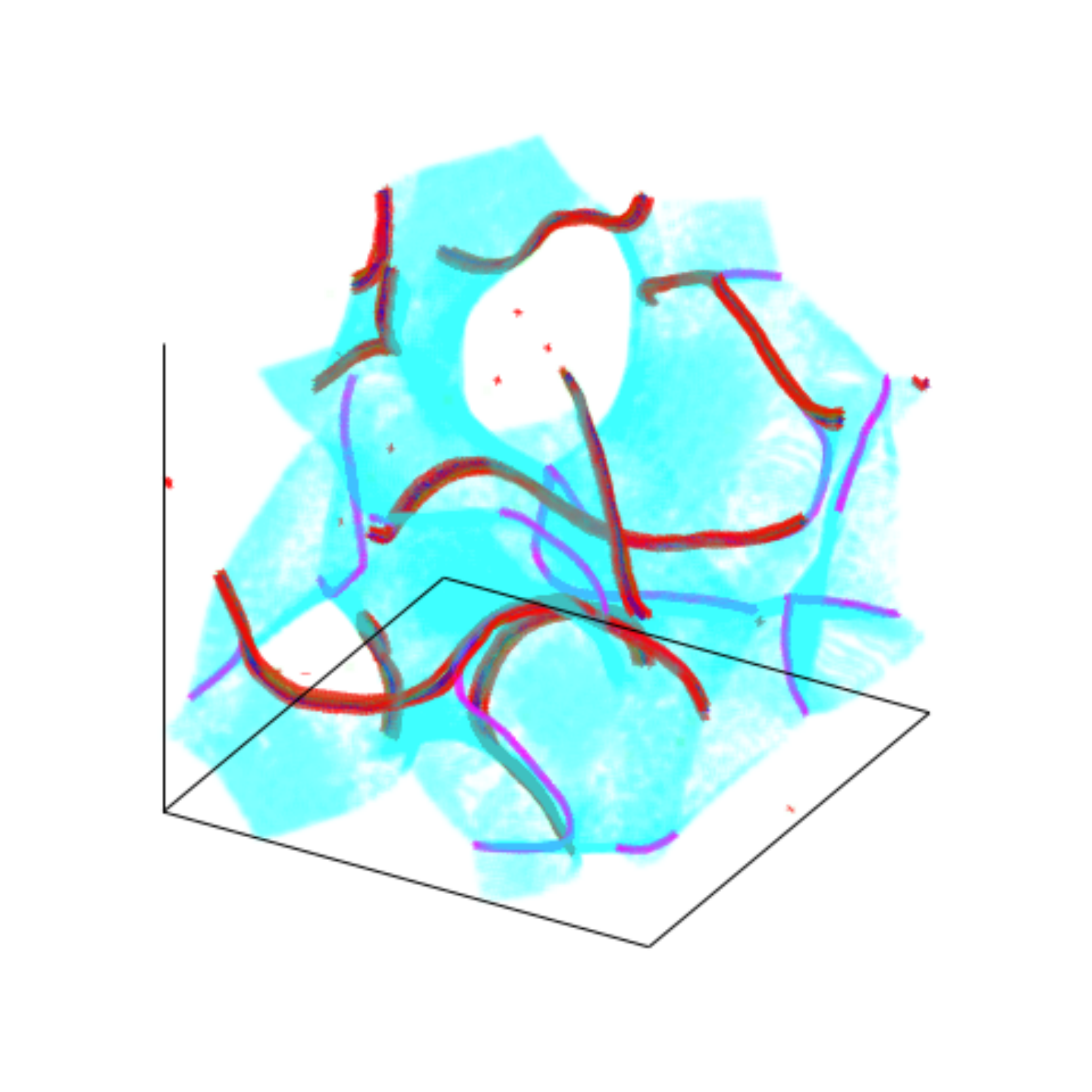}
\label{lattice3D_n3_3}
}
\caption{
The same as Fig.~\ref{lattice_n2} but for $N=3$.
The magenta lines represent $S_3$ and cyan region represents $W_{23}$.
In the right panel, there are a similar number of two kinds of strings, $S_2 + 3 {\bar S}_1$ and $S_3$,
and the walls $W_{23}$ stretching between them.
}
\label{lattice3D_n3}
\end{figure}

We have performed two and three-dimensional lattice simulations for the string-wall network in the aligned axion model
given by Eqs.~(\ref{potential}) and (\ref{clockwork_comp}).
 The results for the two-dimensional simulations with $N=2$ and $3$ are shown in Figs.~\ref{lattice_n2} and \ref{lattice_n3},
 and those for the three-dimensional ones are in  Figs.~\ref{lattice3D_n2} and \ref{lattice3D_n3}.
The initial values of the axions are taken to be random on each grid point.
 
  In the case of $N=2$ (Figs.~\ref{lattice_n2} and \ref{lattice3D_n2}), 
$S_1$ or $\bar{S}_1$ becomes the boundary of domain walls $W_{12}$.
One can see from the figures that such domain walls tend to shrink and eventually (almost) disappear, leading to the cosmic string bundle, 
$S_2 + 3 \bar{S}_1$ or $\bar{S}_2+3S_1$, as discussed before. Thus, isolated string bundles are formed in the case of $N=2$ and $n=3$.

In the case of $N=3$ (Figs.~\ref{lattice_n3} and \ref{lattice3D_n3}), 
the domain wall network survives and the isolated structure cannot be seen. 
In the two-dimensional case, the walls wrap around the lattice simulation box, and both $W_{12}$ and $W_{23}$ remain.
On the other hand, in the three-dimensional case,  while incomplete string bundles, $S_2 + 3 \bar{S}_1$ or $\bar{S}_2+3S_1$, are
formed, there also remain $S_3$ and ${\bar S}_3$ strings. Importantly, their numbers are comparable to each other as expected, which makes
it difficult to form isolated string bundles. We could not find any isolated string bundle and the string-wall system remains
 in the case of $N=n=3$.

The string-wall network likely survives until the QCD phase transition and after that it disappears due to the QCD axion potential which serves as an energy bias to break the degeneracy of discrete vacua.
During the domain wall collisions,  heavy axions are emitted  and they will  decay into gluons.
Depending on the strength of the coupling to gluons and the masses, some of them may be so long-lived that
the energetic gluons change the light element abundances.  In addition to heavy axions, the QCD axion is produced by the 
domain walls and strings. Their abundance may be different from the ordinary QCD axion model, but it requires dedicated numerical
simulations.  The detailed study of those axions produced by the collapse of the string-wall network
is left for future work.

%%%%%%%%%%%%%%% MULTI-FIGURE  %%%%%%%%%%%%%%%
%%%%%%%%%%%%%%%%%%%%%%%%%%%%%%%%%%%%%%%

%%%%%%%%%%%%%%% MULTI-FIGURE  %%%%%%%%%%%%%%%
%%%%%%%%%%%%%%%%%%%%%%%%%%%%%%%%%%%%%%%

%%%%%%%%%%%%%%% MULTI-FIGURE  %%%%%%%%%%%%%%%
%%%%%%%%%%%%%%%%%%%%%%%%%%%%%%%%%%%%%%%

%%%%%%%%%%%%%%% MULTI-FIGURE  %%%%%%%%%%%%%%%
%%%%%%%%%%%%%%%%%%%%%%%%%%%%%%%%%%%%%%%

%%%%%%%%%%%%%%%%%%%%%%%%%%%%%%%%%%%%%%%%%%%%%%%%%%
\subsection{Gravitational waves from domain wall annihilation}
\label{subsec:dw}

We have seen that the domain walls are long-lived for $N \geq 3$
and most probably survive until the QCD axion potential arises.
Assuming the scaling regime, the energy density of domain walls evolves 
as $\rho_{\rm wall} \sim \sigma H$. Here $\sigma$ is the tension of the domain wall,
and it is approximately given by
\begin{align}
\sigma & = 8 m_{aH} f^2,\\
m_{aH} & = {\cal O}(\sqrt{\epsilon} f),
\end{align}
where we have set $f_i = f$ and $\epsilon_i = \epsilon$ for simplicity, and $m_{aH}$ is the mass of heavy axions that form the domain walls. Note that those heavy axions are orthogonal to the QCD axion. 
Since the energy density of domain walls decreases more slowly 
than radiation or matter, stable domain walls will eventually dominate the Universe and make it unacceptably inhomogeneous, spoiling the standard cosmology. 
Thus, domain walls must annihilate before they dominate the Universe. 

In our aligned QCD axion model, the domain walls annihilate after the QCD axion potential arises 
from the QCD instanton effects at $T \sim 1$ GeV 
if the domain wall number of the QCD axion is equal to unity. 
This is because the QCD axion potential behaves like a bias $V_{\epsilon} \sim \Lambda_{\rm QCD}^4$ for the  $W_{N-1,N}$ domain walls and they annihilate at $\sigma H \sim \Lambda_{\rm QCD}^4$ \cite{Gelmini:1988sf}. Remaining domain walls $W_{N-2,N-1},\dots, W_{1,2}$ also annihilate one after another with attached cosmic strings.
If $\sigma H < \Lambda_{\rm QCD}^4$ is satisfied at $T \sim 1$ GeV, the domain walls annihilate soon after the QCD axion potential is turned on.
Otherwise, the annihilation temperature is estimated as
\beq
T_{\rm ann} \sim 1~{\rm GeV}~\epsilon^{-1/4}\bigg(\frac{g_*(T_{\rm ann})}{80}\bigg)^{-1/4} \bigg(\frac{\Lambda_{\rm QCD}}{400~{\rm MeV}} \bigg)^2 \bigg( \frac{100~{\rm TeV}}{f} \bigg)^{3/2},
\eeq
which implies that the annihilation is delayed significantly for $f \gtrsim 100~{\rm TeV}~\epsilon^{1/6}$.
We have numerically confirmed that the string-wall system collapses after the QCD instanton effects become relevant.
Requiring that the domain walls should not dominate the Universe until $T=T_{\rm ann}$, one obtains a conservative constraint on $f$,
\beq
f \lesssim 400~{\rm TeV}~\epsilon^{-1/6} \bigg( \frac{\Lambda_{\rm QCD}}{400~{\rm MeV}} \bigg)^{4/3},
\eeq
and the annihilation temperature must be $T_{\rm ann} \gtrsim 0.1$ GeV.\footnote{%%
In the presence of extra PQ breaking terms, each axion is trapped at each potential minima even before the QCD phase transition \cite{Higaki:2016yqk}.
In this case, the string-wall network sooner disappears and $f$ can be a larger value.
}%%
In fact, a more stringent constraint comes from the  pulsar timing observation, as we shall see below.

In the violent collisions of domain walls, gravitational waves are produced over frequencies $\nu$  corresponding to a typical physical length scale~\cite{Gleiser:1998na,Hiramatsu:2010yz,Kawasaki:2011vv,Hiramatsu:2013qaa}. Once the domain-wall network follows the scaling law, 
a typical curvature radius is of order the Hubble parameter. Then, 
the power spectrum of the gravitational waves has a peak at frequency corresponding to the Hubble parameter at the domain wall annihilation, $\nu_{\rm peak}(t_{\rm ann}) \simeq H_{\rm ann}$. As the Universe expands, the peak frequency is red-shifted, and
its present value is given by
\beq
\nu_{\rm peak,0} \simeq 1.6 \times 10^{-7}~{\rm Hz}~\bigg(\frac{g_{*{\rm ann}}}{80}\bigg)^{1/6} \bigg(\frac{T_{\rm ann}}{1~{\rm GeV}} \bigg),
\eeq
which happens to be in the sensitivity range of the pulsar timing experiments.

The intensity of the gravitational waves is usually characterized by a dimensionless quantity, $\Omega_{\rm gw}(t)$,
defined by
\begin{equation}
    \Omega_{\rm gw}(\nu(t)) = \frac{1}{\rho_c(t)} 
    \frac{d \log \rho_{\rm gw}(t)}{d \log \nu},
\end{equation}
where  $\rho_{\rm gw}(t)$ is the energy density of gravitational waves,
$\rho_c(t)$  the critical density, and $\nu$ is the frequency.
The peak value of $\Omega_{\rm gw}$ at the annihilation is estimated as
\beq
\Omega_{\rm gw} (\nu_{\rm peak}(t_{\rm ann}))= 
%\left.
\frac{8 \pi \tilde{\epsilon}_{\rm gw} G^2 \mathcal{A}^2 \sigma^2}{3 H_{\rm ann} ^2},
%\right|_{t = t_{\rm ann}}
\eeq
where $\tilde{\epsilon}_{\rm gw} \simeq 0.7 \pm 0.4$~\cite{Hiramatsu:2013qaa} is an efficiency parameter of the gravitational wave emission,
and $\mathcal{A}$ parametrizes the energy density of domain walls as $\rho_{\rm wall} = \mathcal{A} \sigma /t $
in the radiation-dominated era. In the case of a simple $Z_2$ potential, $\mathcal{A} \simeq 0.8 \pm 0.1$, and it is considered to increase 
in proportion to the number of the potential minima~\cite{Hiramatsu:2012sc,Kawasaki:2014sqa}. In our aligned axion model, the vacuum
structure is much more complicated than the case of $Z_2$, and we expect $\mathcal{A} = {\cal O}(1-10)$.
In fact, we have obtained ${\cal A} \sim N$ from two dimensional lattice simulations.

The present value of $\Omega_{\rm gw}$ at the peak frequency is given by
\begin{align}
\Omega_{\rm gw}(\nu_{\rm peak}(t_0)) h^2 &= \Omega_R h^2 \left(
\frac{g_{*s,0}^\frac{4}{3}/g_{*,0}}{g_{*{\rm ann}}^{1/3}}
\right)
\Omega_{\rm gw} (\nu_{\rm peak}(t_{\rm ann})) \nonumber \\
&\simeq 
2 \times 10^{-11} \epsilon \lrfp{g_{*{\rm ann}}}{80}{-\frac{4}{3}}\lrf{\tilde{\epsilon}_{\rm gw}}{0.7}\lrfp{{\cal A}}{10}{2} 
 \lrfp{T_{\rm ann}}{1{\rm\,GeV}}{-4} 
  \lrfp{f}{10^2 {\rm\,TeV}}{6},
\end{align}
where $\Omega_R h^2 \simeq 4.15 \times 10^{-5}$ is the density parameter of radiation (assuming massless neutrinos), $g_{*,0} \simeq 3.363$, $g_{*s,0} \simeq 3.909$, and $g_{* {\rm ann}}$ are the effective relativistic degrees of freedom. The subscript `$0$' and `ann' represent that the variable is evaluated at present and at the domain wall annihilation, respectively.

Interestingly, the peak frequency with $T_{\rm ann} \sim 1$ GeV is within the target range of pulsar timing observations \cite{Arzoumanian:2015liz,Lentati:2015qwp,Verbiest:2016vem,Lasky:2015lej}. The current $95\%$ confidence upper limit reads  
$\Omega_{\rm gw} h^2 < 2.3 \times 10^{-10}$ at  $\nu_{\rm 1yr} \simeq 3\times 10^{-8}$~Hz~\cite{Lasky:2015lej}.
The gravitational waves from the domain wall annihilation have a frequency dependence, $\Omega_{\rm gw} \propto \nu^3$, at $\nu < \nu_{\rm peak}$.
Thus we obtain the upper bound on $f$,
\beq
f \lesssim 200~{\rm TeV} \times \epsilon^{-\frac{1}{6}} \bigg(\frac{g_{*{\rm ann}}}{80}\bigg)^{\frac{1}{198}}\lrfp{\epsilon_{\rm gw}}{0.7}{-\frac{2}{33}}\lrfp{{\cal A}}{10}{-\frac{4}{33} }
\lrfp{\Lambda_{\rm QCD}}{400~{\rm MeV}}{\frac{28}{33}}.
\eeq
As mentioned before, it may take a longer time for the string-wall system to annihilate completely in the aligned axion model,
compared to the usual case of a single PQ scalar. As a result, the gravitational power spectrum may be modified, and, in particular, the peak frequency
may be slightly lowered. Taking account of this uncertainty, the upper bound on $f$ may be tightened by a factor of several.
Detailed lattice numerical simulations for the produced gravitaional waves are warranted.

The future observation by SKA will reach $\Omega_{\rm gw} h^2 \sim 10^{-13}$ \cite{Janssen:2014dka} and the upper bound on $f$
will be improved by a factor of $2$.
It implies that the gravitational waves produced in the aligned axion model with a decay constant of ${\cal O}(10-100)$ TeV 
will be probed by the pulsar timing experiment in future.

%%%%%%%%%%%%%%%%%%%%%%%%%%%%%%%%%%%%%%
\section{Discussion and conclusions}
\label{sec:conc}
%%%%%%%%%%%%%%%%%%%%%%%%%%%%%%%%%%%%%%

Let us discuss the case in which the U(1)$^N$ symmetries are only partially restored by
some of the scalar fields which develop VEVs after inflation.
Suppose that $\Phi_N$ develops a non-zero VEV during inflation.
In this case, a complicated string-wall network does not remain for a long time. This is because, for the cosmic string to remain in the end,
the phase of $\Phi_N$ should rotate from 0 to 2$\pi$ about the string, but if $\Phi_N$ has
a non-zero VEV already, such configuration does not appear.
It seems that only the cosmic strings associated with the scalars which develop non-zero 
VEVs after inflation will appear for the moment, but they are attached to domain walls and so, they will disappear 
as soon as the domain walls appear.
As for the isocurvature fluctuations, only some fraction of the field space
is populated by this partial spontaneous symmetry breaking. Therefore, after the string-wall network disappears, 
some amount of isocurvature perturbations may be left.  But it is difficult to know what actually happens 
unless the evolution of the string-wall network is numerically studied.  In any case, complete symmetry restoration seems more plausible, in which case
no isocurvature perturbations are generated. 

In this paper we have studied the formation and evolution of the complicated string-wall network
in the aligned axion model. Since the actual PQ symmetry breaking scale is much smaller than 
the classical axion window, the symmetry restoration is more plausible in this scenario.
Such axion model has several virtues: the high quality of the PQ symmetry is naturally explained,
and no isocurvature perturbations are generated if the symmetry is restored during or after inflation.
We have shown that 
there exists a solution of the isolated string-wall system, which can be identified as the QCD axion string
as their string tensions are approximately equal to each other.
However,  the formation probability of such isolated string bundles  is extremely suppressed for $N \geq 3$,
because it requires a highly biased population of each type of strings ($S_i$) over anti-strings ($\bar{S}_i$),
which cannot be realized if the strings follow the scaling law.
Thus, the string-wall network is considered to be extremely long-lived, and most probably survives until the QCD axion potential appears.
During the violent domain wall collisions, a significant amount of gravitational waves are emitted. We have derived
an upper bound on the axion decay constant, $f_i \lesssim {\cal O}(100)$\,TeV, by the pulsar timing experiments,
and the bound will be improved by a factor of $2$ in the future observations by SKA.

%%%%%%%%%%%%%%%%%%%%%%%%%%%%%%%%%%%%%&
\section*{Acknowledgment}
%%%%%%%%%%%%%%%%%%%%%%%%%%%%%%%%%%%%%&

This work is supported by MEXT KAKENHI Grant Numbers 15H05889 and 15K21733 (F.T.),
JSPS KAKENHI Grant Numbers  26247042(F.T. and T.H.), and 26287039 (F.T.),
World Premier International Research Center Initiative (WPI Initiative), MEXT, Japan (F.T.),
and MEXT-Supported Program for the Strategic Research Foundation at Private Universities,
"Topological Science", Grant Number S1511006 (T.H.),
the Max-Planck-Gesellschaft, the Korea Ministry of Education, Science and Technology,
Gyeongsangbuk-Do and Pohang City for the support of the Independent Junior Research Group at the
Asia Pacific Center for Theoretical Physics (N.K.).
This work is also supported by the National Research Foundation of Korea (NRF) grant funded by the Korea government (MSIP)
(NRF-2015R1D1A3A01019746) (K.S.J).
We would like to thank IBS for the computational resource and financial support under the project code IBS-R018-D1 (T.S.).


\begin{thebibliography}{99}

%\cite{Peccei:1977hh}
\bibitem{Peccei:1977hh} 
  R.~D.~Peccei and H.~R.~Quinn,
  %``CP Conservation in the Presence of Instantons,''
  Phys.\ Rev.\ Lett.\  {\bf 38}, 1440 (1977).
  %doi:10.1103/PhysRevLett.38.1440
  %%CITATION = doi:10.1103/PhysRevLett.38.1440;%%
  %3512 citations counted in INSPIRE as of 28 Apr 2016

%\cite{Peccei:1977ur}
\bibitem{Peccei:1977ur} 
  R.~D.~Peccei and H.~R.~Quinn,
  %``Constraints Imposed by CP Conservation in the Presence of Instantons,''
  Phys.\ Rev.\ D {\bf 16}, 1791 (1977).
  %doi:10.1103/PhysRevD.16.1791
  %%CITATION = doi:10.1103/PhysRevD.16.1791;%%
  %2032 citations counted in INSPIRE as of 28 Apr 2016
  
  %\cite{Weinberg:1977ma}
\bibitem{Weinberg:1977ma} 
  S.~Weinberg,
  %``A New Light Boson?,''
  Phys.\ Rev.\ Lett.\  {\bf 40}, 223 (1978).
  %doi:10.1103/PhysRevLett.40.223
  %%CITATION = doi:10.1103/PhysRevLett.40.223;%%
  %2460 citations counted in INSPIRE as of 28 Apr 2016

%\cite{Wilczek:1977pj}
\bibitem{Wilczek:1977pj} 
  F.~Wilczek,
  %``Problem of Strong p and t Invariance in the Presence of Instantons,''
  Phys.\ Rev.\ Lett.\  {\bf 40}, 279 (1978).
  %doi:10.1103/PhysRevLett.40.279
  %%CITATION = doi:10.1103/PhysRevLett.40.279;%%
  %2375 citations counted in INSPIRE as of 28 Apr 2016
  
  
  %\cite{Mayle:1987as}
\bibitem{Mayle:1987as} 
  R.~Mayle, J.~R.~Wilson, J.~R.~Ellis, K.~A.~Olive, D.~N.~Schramm and G.~Steigman,
  %``Constraints on Axions from SN 1987a,''
  Phys.\ Lett.\ B {\bf 203}, 188 (1988).
%  doi:10.1016/0370-2693(88)91595-X
  %%CITATION = doi:10.1016/0370-2693(88)91595-X;%%
  %144 citations counted in INSPIRE as of 05 f\UTF{00E9}vr. 2016

%\cite{Raffelt:1987yt}
\bibitem{Raffelt:1987yt} 
  G.~Raffelt and D.~Seckel,
  %``Bounds on Exotic Particle Interactions from SN 1987a,''
  Phys.\ Rev.\ Lett.\  {\bf 60}, 1793 (1988).
%  doi:10.1103/PhysRevLett.60.1793
  %%CITATION = doi:10.1103/PhysRevLett.60.1793;%%
  %311 citations counted in INSPIRE as of 05 f\UTF{00E9}vr. 2016
  
  %\cite{Turner:1987by}
\bibitem{Turner:1987by} 
  M.~S.~Turner,
  %``Axions from SN 1987a,''
  Phys.\ Rev.\ Lett.\  {\bf 60}, 1797 (1988).
  %doi:10.1103/PhysRevLett.60.1797
  %%CITATION = doi:10.1103/PhysRevLett.60.1797;%%
  %231 citations counted in INSPIRE as of 05 f\UTF{00E9}vr. 2016
  
  
%\cite{Preskill:1982cy}
\bibitem{Preskill:1982cy} 
  J.~Preskill, M.~B.~Wise and F.~Wilczek,
  %``Cosmology of the Invisible Axion,''
  Phys.\ Lett.\ B {\bf 120}, 127 (1983).
%  doi:10.1016/0370-2693(83)90637-8
  %%CITATION = doi:10.1016/0370-2693(83)90637-8;%%
  %1174 citations counted in INSPIRE as of 04 May 2016
  
  %\cite{Abbott:1982af}
\bibitem{Abbott:1982af} 
  L.~F.~Abbott and P.~Sikivie,
  %``A Cosmological Bound on the Invisible Axion,''
  Phys.\ Lett.\ B {\bf 120}, 133 (1983).
%  doi:10.1016/0370-2693(83)90638-X
  %%CITATION = doi:10.1016/0370-2693(83)90638-X;%%
  %1142 citations counted in INSPIRE as of 04 May 2016
  
  %\cite{Dine:1982ah}
\bibitem{Dine:1982ah} 
  M.~Dine and W.~Fischler,
  %``The Not So Harmless Axion,''
  Phys.\ Lett.\ B {\bf 120}, 137 (1983).
%  doi:10.1016/0370-2693(83)90639-1
  %%CITATION = doi:10.1016/0370-2693(83)90639-1;%%
  %1112 citations counted in INSPIRE as of 04 May 2016
  

  %\cite{Kim:1986ax}
\bibitem{Kim:1986ax} 
  J.~E.~Kim,
  %``Light Pseudoscalars, Particle Physics and Cosmology,''
  Phys.\ Rept.\  {\bf 150}, 1 (1987).
  %%CITATION = PRPLC,150,1;%%
  %851 citations counted in INSPIRE as of 24 sept. 2015

  
%\cite{Kim:2008hd}
\bibitem{Kim:2008hd} 
  J.~E.~Kim and G.~Carosi,
  %``Axions and the Strong CP Problem,''
  Rev.\ Mod.\ Phys.\  {\bf 82}, 557 (2010)
  [arXiv:0807.3125 [hep-ph]].
  %%CITATION = ARXIV:0807.3125;%%
  %293 citations counted in INSPIRE as of 24 sept. 2015
  
  %\cite{Wantz:2009it}
\bibitem{Wantz:2009it} 
  O.~Wantz and E.~P.~S.~Shellard,
  %``Axion Cosmology Revisited,''
  Phys.\ Rev.\ D {\bf 82}, 123508 (2010)
  [arXiv:0910.1066 [astro-ph.CO]].
  %%CITATION = ARXIV:0910.1066;%%
  %95 citations counted in INSPIRE as of 24 sept. 2015
  
  %\cite{Ringwald:2012hr}
\bibitem{Ringwald:2012hr}
 A.~Ringwald,
 %``Exploring the Role of Axions and Other WISPs in the Dark Universe,''
 Phys.\ Dark Univ.\  {\bf 1} (2012) 116.
 %[arXiv:1210.5081 [hep-ph]];
 %%CITATION = ARXIV:1210.5081;%%
 
%\cite{Kawasaki:2013ae}
\bibitem{Kawasaki:2013ae} 
  M.~Kawasaki and K.~Nakayama,
  %``Axions: Theory and Cosmological Role,''
  Ann.\ Rev.\ Nucl.\ Part.\ Sci.\  {\bf 63}, 69 (2013)
  [arXiv:1301.1123 [hep-ph]].
  %%CITATION = ARXIV:1301.1123;%%
  %60 citations counted in INSPIRE as of 22 sept. 2015

  %\cite{Conlon:2006tq}
\bibitem{Conlon:2006tq} 
  J.~P.~Conlon,
  %``The QCD axion and moduli stabilisation,''
  JHEP {\bf 0605}, 078 (2006)
  %doi:10.1088/1126-6708/2006/05/078
  [hep-th/0602233].
  %%CITATION = doi:10.1088/1126-6708/2006/05/078;%%
  %134 citations counted in INSPIRE as of 16 Jun 2016
  
  %\cite{Svrcek:2006yi}
\bibitem{Svrcek:2006yi} 
  P.~Svrcek and E.~Witten,
  %``Axions In String Theory,''
  JHEP {\bf 0606}, 051 (2006)
  %doi:10.1088/1126-6708/2006/06/051
  [hep-th/0605206].
  %%CITATION = doi:10.1088/1126-6708/2006/06/051;%%
  %377 citations counted in INSPIRE as of 28 Apr 2016
  
  %\cite{Arvanitaki:2009fg}
\bibitem{Arvanitaki:2009fg} 
  A.~Arvanitaki, S.~Dimopoulos, S.~Dubovsky, N.~Kaloper and J.~March-Russell,
  %``String Axiverse,''
  Phys.\ Rev.\ D {\bf 81}, 123530 (2010)
  %doi:10.1103/PhysRevD.81.123530
  [arXiv:0905.4720 [hep-th]].
  %%CITATION = doi:10.1103/PhysRevD.81.123530;%%
  %278 citations counted in INSPIRE as of 28 Apr 2016
  
  %\cite{Cicoli:2012sz}
\bibitem{Cicoli:2012sz} 
  M.~Cicoli, M.~Goodsell and A.~Ringwald,
  %``The type IIB string axiverse and its low-energy phenomenology,''
  JHEP {\bf 1210}, 146 (2012)
  %doi:10.1007/JHEP10(2012)146
  [arXiv:1206.0819 [hep-th]].
  %%CITATION = doi:10.1007/JHEP10(2012)146;%%
  %95 citations counted in INSPIRE as of 28 Apr 2016
  
  
  
  %\cite{Kim:2004rp}
\bibitem{Kim:2004rp} 
  J.~E.~Kim, H.~P.~Nilles and M.~Peloso,
  %``Completing natural inflation,''
  JCAP {\bf 0501}, 005 (2005)
  %doi:10.1088/1475-7516/2005/01/005
  [hep-ph/0409138].
  %%CITATION = doi:10.1088/1475-7516/2005/01/005;%%
  %235 citations counted in INSPIRE as of 28 Apr 2016
  


%\cite{Choi:2014rja}
\bibitem{Choi:2014rja} 
  K.~Choi, H.~Kim and S.~Yun,
  %``Natural inflation with multiple sub-Planckian axions,''
  Phys.\ Rev.\ D {\bf 90}, 023545 (2014)
  %doi:10.1103/PhysRevD.90.023545
  [arXiv:1404.6209 [hep-th]].
  %%CITATION = doi:10.1103/PhysRevD.90.023545;%%
  %68 citations counted in INSPIRE as of 28 Apr 2016

%\cite{Higaki:2014pja}
\bibitem{Higaki:2014pja} 
  T.~Higaki and F.~Takahashi,
  %``Natural and Multi-Natural Inflation in Axion Landscape,''
  JHEP {\bf 1407}, 074 (2014)
  doi:10.1007/JHEP07(2014)074
  [arXiv:1404.6923 [hep-th]].
  %%CITATION = doi:10.1007/JHEP07(2014)074;%%
  %44 citations counted in INSPIRE as of 29 Apr 2016
  
  %\cite{Higaki:2014mwa}
\bibitem{Higaki:2014mwa} 
  T.~Higaki and F.~Takahashi,
  %``Axion Landscape and Natural Inflation,''
  Phys.\ Lett.\ B {\bf 744}, 153 (2015)
  %doi:10.1016/j.physletb.2015.03.052
  [arXiv:1409.8409 [hep-ph]].
  %%CITATION = doi:10.1016/j.physletb.2015.03.052;%%
  %20 citations counted in INSPIRE as of 29 Apr 2016

%\cite{Kappl:2014lra}
\bibitem{Kappl:2014lra} 
  R.~Kappl, S.~Krippendorf and H.~P.~Nilles,
  %``Aligned Natural Inflation: Monodromies of two Axions,''
  Phys.\ Lett.\ B {\bf 737}, 124 (2014)
  %doi:10.1016/j.physletb.2014.08.045
  [arXiv:1404.7127 [hep-th]].
  %%CITATION = doi:10.1016/j.physletb.2014.08.045;%%
  %74 citations counted in INSPIRE as of 28 Apr 2016
  
  %\cite{Ben-Dayan:2014zsa}
\bibitem{Ben-Dayan:2014zsa} 
  I.~Ben-Dayan, F.~G.~Pedro and A.~Westphal,
  %``Hierarchical Axion Inflation,''
  Phys.\ Rev.\ Lett.\  {\bf 113}, 261301 (2014)
 % doi:10.1103/PhysRevLett.113.261301
  [arXiv:1404.7773 [hep-th]].
  %%CITATION = doi:10.1103/PhysRevLett.113.261301;%%
  %51 citations counted in INSPIRE as of 19 May 2016
  
  %\cite{Long:2014dta}
\bibitem{Long:2014dta} 
  C.~Long, L.~McAllister and P.~McGuirk,
  %``Aligned Natural Inflation in String Theory,''
  Phys.\ Rev.\ D {\bf 90}, 023501 (2014)
  %doi:10.1103/PhysRevD.90.023501
  [arXiv:1404.7852 [hep-th]].
  %%CITATION = doi:10.1103/PhysRevD.90.023501;%%
  %61 citations counted in INSPIRE as of 16 Jun 2016
  
  %\cite{Harigaya:2014rga}
\bibitem{Harigaya:2014rga} 
  K.~Harigaya and M.~Ibe,
  %``Phase Locked Inflation -- Effectively Trans-Planckian Natural Inflation,''
  JHEP {\bf 1411}, 147 (2014)
  %doi:10.1007/JHEP11(2014)147
  [arXiv:1407.4893 [hep-ph]].
  %%CITATION = doi:10.1007/JHEP11(2014)147;%%
  %8 citations counted in INSPIRE as of 03 May 2016
  
  %\cite{Choi:2015fiu}\cite{Kaplan:2015fuy}
\bibitem{Choi:2015fiu} 
  K.~Choi and S.~H.~Im,
  %``Realizing the relaxion from multiple axions and its UV completion with high scale supersymmetry,''
  JHEP {\bf 1601}, 149 (2016)
  %doi:10.1007/JHEP01(2016)149
  [arXiv:1511.00132 [hep-ph]].
  %%CITATION = doi:10.1007/JHEP01(2016)149;%%
  %13 citations counted in INSPIRE as of 28 Apr 2016
  
  %\cite{Kaplan:2015fuy}
\bibitem{Kaplan:2015fuy} 
  D.~E.~Kaplan and R.~Rattazzi,
  %``Large field excursions and approximate discrete symmetries from a clockwork axion,''
  Phys.\ Rev.\ D {\bf 93}, no. 8, 085007 (2016)
  %doi:10.1103/PhysRevD.93.085007
  [arXiv:1511.01827 [hep-ph]].
  %%CITATION = doi:10.1103/PhysRevD.93.085007;%%
  %14 citations counted in INSPIRE as of 28 Apr 2016


  
  %\cite{Kobayashi:2010pz}
\bibitem{Kobayashi:2010pz} 
  T.~Kobayashi and F.~Takahashi,
  %``Running Spectral Index from Inflation with Modulations,''
  JCAP {\bf 1101}, 026 (2011)
  %doi:10.1088/1475-7516/2011/01/026
  [arXiv:1011.3988 [astro-ph.CO]].
  %%CITATION = doi:10.1088/1475-7516/2011/01/026;%%
  %54 citations counted in INSPIRE as of 20 May 2016
  
  %\cite{Czerny:2014wua}
\bibitem{Czerny:2014wua} 
  M.~Czerny, T.~Kobayashi and F.~Takahashi,
  %``Running Spectral Index from Large-field Inflation with Modulations Revisited,''
  Phys.\ Lett.\ B {\bf 735}, 176 (2014)
  %doi:10.1016/j.physletb.2014.06.018
  [arXiv:1403.4589 [astro-ph.CO]].
  %%CITATION = doi:10.1016/j.physletb.2014.06.018;%%
  %44 citations counted in INSPIRE as of 26 May 2016
  
  %\cite{Wang:2015rel}
\bibitem{Wang:2015rel} 
  G.~Wang and T.~Battefeld,
  %``Vacuum Selection on Axionic Landscapes,''
  JCAP {\bf 1604}, no. 04, 025 (2016)
  %doi:10.1088/1475-7516/2016/04/025
  [arXiv:1512.04224 [hep-th]].
  %%CITATION = doi:10.1088/1475-7516/2016/04/025;%%
  %3 citations counted in INSPIRE as of 19 May 2016

%\cite{Masoumi:2016eqo}
\bibitem{Masoumi:2016eqo} 
  A.~Masoumi and A.~Vilenkin,
  %``Vacuum statistics and stability in axionic landscapes,''
  JCAP {\bf 1603}, no. 03, 054 (2016)
  %doi:10.1088/1475-7516/2016/03/054
  [arXiv:1601.01662 [gr-qc]].
  %%CITATION = doi:10.1088/1475-7516/2016/03/054;%%
  %2 citations counted in INSPIRE as of 19 May 2016


%\cite{Higaki:2015jag}
\bibitem{Higaki:2015jag} 
  T.~Higaki, K.~S.~Jeong, N.~Kitajima and F.~Takahashi,
  %``The QCD Axion from Aligned Axions and Diphoton Excess,''
  Phys.\ Lett.\ B {\bf 755}, 13 (2016)
  %doi:10.1016/j.physletb.2016.01.055
  [arXiv:1512.05295 [hep-ph]].
  %%CITATION = doi:10.1016/j.physletb.2016.01.055;%%
  %127 citations counted in INSPIRE as of 28 Apr 2016
  
  %\cite{Higaki:2016yqk}
\bibitem{Higaki:2016yqk} 
  T.~Higaki, K.~S.~Jeong, N.~Kitajima and F.~Takahashi,
  %``Quality of the Peccei-Quinn symmetry in the Aligned QCD Axion and Cosmological Implications,''
  arXiv:1603.02090 [hep-ph].
  %%CITATION = ARXIV:1603.02090;%%
  %2 citations counted in INSPIRE as of 28 Apr 2016
  
  %\cite{}
\bibitem{ATLAS} 
  The ATLAS collaboration,
  %``Search for resonances decaying to photon pairs in 3.2 fb$^{-1}$ of $pp$ collisions at $\sqrt{s}$ = 13 TeV with the ATLAS detector,''
  ATLAS-CONF-2015-081;
  %%CITATION = ATLAS-CONF-2015-081;%%
  %282 citations counted in INSPIRE as of 07 Apr 2016
%
%%\cite{}
%\bibitem{ATALS2} 
%  The ATLAS collaboration,
  %``Search for resonances in diphoton events with the ATLAS detector at $\sqrt{s}$ = 13 TeV,''
  ATLAS-CONF-2016-018.
  %%CITATION = ATLAS-CONF-2016-018;%%
  %2 citations counted in INSPIRE as of 07 Apr 2016
  
%\cite{CMS:2015dxe}
\bibitem{CMS:2015dxe} 
  CMS Collaboration [CMS Collaboration],
  %``Search for new physics in high mass diphoton events in proton-proton  collisions at 13TeV,''
  CMS-PAS-EXO-15-004;
  %%CITATION = CMS-PAS-EXO-15-004;%%
  %272 citations counted in INSPIRE as of 07 Apr 2016
%\cite{CMS:2016owr}
%\bibitem{CMS:2016owr} 
%  CMS Collaboration [CMS Collaboration],
  %``Search for new physics in high mass diphoton events in $3.3~\mathrm{fb}^{-1}$ of proton-proton collisions at $\sqrt{s}=13~\mathrm{TeV}$ and combined interpretation of searches at $8~\mathrm{TeV}$ and $13~\mathrm{TeV}$,''
  CMS-PAS-EXO-16-018.
  %%CITATION = CMS-PAS-EXO-16-018;%%
  %8 citations counted in INSPIRE as of 07 Apr 2016

%\cite{Chiang:2016eav}
\bibitem{Chiang:2016eav} 
 C.~W.~Chiang, H.~Fukuda, M.~Ibe and T.~T.~Yanagida,
 %``750 GeV diphoton resonance in a visible heavy QCD axion model,''
 Phys.\ Rev.\ D {\bf 93}, no. 9, 095016 (2016)
 %doi:10.1103/PhysRevD.93.095016
 [arXiv:1602.07909 [hep-ph]].
 %%CITATION = doi:10.1103/PhysRevD.93.095016;%%
 %15 citations counted in INSPIRE as of 13 Jun 2016

%\cite{Gherghetta:2016fhp}
\bibitem{Gherghetta:2016fhp} 
 T.~Gherghetta, N.~Nagata and M.~Shifman,
 %``A Visible QCD Axion from an Enlarged Color Group,''
 Phys.\ Rev.\ D {\bf 93}, no. 11, 115010 (2016)
 %doi:10.1103/PhysRevD.93.115010
 [arXiv:1604.01127 [hep-ph]].
 %%CITATION = doi:10.1103/PhysRevD.93.115010;%%
 %6 citations counted in INSPIRE as of 13 Jun 2016

%\cite{DHHM}
\bibitem{DHHM} 
 S.~Dimopoulos, A.~Hook, J.~Huang and G.~Marques-Tavares,
 %``A 750 GeV QCD axion,''
 arXiv:1606.03097 [hep-ph].
 %%CITATION = ARXIV:1606.03097;%%

%\cite{Carpenter:2009zs}
\bibitem{Carpenter:2009zs} 
  L.~M.~Carpenter, M.~Dine and G.~Festuccia,
  %``Dynamics of the Peccei Quinn Scale,''
  Phys.\ Rev.\ D {\bf 80}, 125017 (2009)
  %doi:10.1103/PhysRevD.80.125017
  [arXiv:0906.1273 [hep-th]].
  %%CITATION = doi:10.1103/PhysRevD.80.125017;%%
  %22 citations counted in INSPIRE as of 28 Apr 2016

%\cite{Fukuda:2015ana}
\bibitem{Fukuda:2015ana} 
 H.~Fukuda, K.~Harigaya, M.~Ibe and T.~T.~Yanagida,
 %``Model of visible QCD axion,''
 Phys.\ Rev.\ D {\bf 92}, no. 1, 015021 (2015)
 %doi:10.1103/PhysRevD.92.015021
 [arXiv:1504.06084 [hep-ph]].
 %%CITATION = doi:10.1103/PhysRevD.92.015021;%%
 %7 citations counted in INSPIRE as of 13 Jun 2016

%\cite{Hiramatsu:2012sc}
\bibitem{Hiramatsu:2012sc} 
  T.~Hiramatsu, M.~Kawasaki, K.~Saikawa and T.~Sekiguchi,
  %``Axion cosmology with long-lived domain walls,''
  JCAP {\bf 1301}, 001 (2013)
  %doi:10.1088/1475-7516/2013/01/001
  [arXiv:1207.3166 [hep-ph]].
  %%CITATION = doi:10.1088/1475-7516/2013/01/001;%%
  %29 citations counted in INSPIRE as of 23 May 2016

%\cite{Kawasaki:2014sqa}
\bibitem{Kawasaki:2014sqa} 
  M.~Kawasaki, K.~Saikawa and T.~Sekiguchi,
  %``Axion dark matter from topological defects,''
  Phys.\ Rev.\ D {\bf 91}, no. 6, 065014 (2015)
  %doi:10.1103/PhysRevD.91.065014
  [arXiv:1412.0789 [hep-ph]].
  %%CITATION = doi:10.1103/PhysRevD.91.065014;%%
  %19 citations counted in INSPIRE as of 28 Apr 2016




%\cite{Daido:2015bva}
\bibitem{Daido:2015bva} 
  R.~Daido, N.~Kitajima and F.~Takahashi,
  %``Domain Wall Formation from Level Crossing in the Axiverse,''
  Phys.\ Rev.\ D {\bf 92}, no. 6, 063512 (2015)
  %doi:10.1103/PhysRevD.92.063512
  [arXiv:1505.07670 [hep-ph]].
  %%CITATION = doi:10.1103/PhysRevD.92.063512;%%
  %2 citations counted in INSPIRE as of 28 Apr 2016

%\cite{Daido:2015cba}
\bibitem{Daido:2015cba} 
  R.~Daido, N.~Kitajima and F.~Takahashi,
  %``Level crossing between the QCD axion and an axionlike particle,''
  Phys.\ Rev.\ D {\bf 93}, no. 7, 075027 (2016)
  %doi:10.1103/PhysRevD.93.075027
  [arXiv:1510.06675 [hep-ph]].
  %%CITATION = doi:10.1103/PhysRevD.93.075027;%%
  %1 citations counted in INSPIRE as of 28 Apr 2016
  
     %\cite{Lentati:2015qwp}
\bibitem{Lentati:2015qwp} 
  L.~Lentati {\it et al.},
  %``European Pulsar Timing Array Limits On An Isotropic Stochastic Gravitational-Wave Background,''
  Mon.\ Not.\ Roy.\ Astron.\ Soc.\  {\bf 453}, no. 3, 2576 (2015)
  %doi:10.1093/mnras/stv1538
  [arXiv:1504.03692 [astro-ph.CO]].
  %%CITATION = doi:10.1093/mnras/stv1538;%%
  %38 citations counted in INSPIRE as of 16 May 2016
  
  %\cite{Arzoumanian:2015liz}
\bibitem{Arzoumanian:2015liz} 
  Z.~Arzoumanian {\it et al.} [NANOGrav Collaboration],
  %``The NANOGrav Nine-year Data Set: Limits on the Isotropic Stochastic Gravitational Wave Background,''
  Astrophys.\ J.\  {\bf 821}, no. 1, 13 (2016)
  %doi:10.3847/0004-637X/821/1/13
  [arXiv:1508.03024 [astro-ph.GA]].
  %%CITATION = doi:10.3847/0004-637X/821/1/13;%%
  %25 citations counted in INSPIRE as of 16 May 2016
  
  %\cite{Lasky:2015lej}
\bibitem{Lasky:2015lej} 
  P.~D.~Lasky {\it et al.},
  %``Gravitational-wave cosmology across 29 decades in frequency,''
  Phys.\ Rev.\ X {\bf 6}, no. 1, 011035 (2016)
  %doi:10.1103/PhysRevX.6.011035
  [arXiv:1511.05994 [astro-ph.CO]].
  %%CITATION = doi:10.1103/PhysRevX.6.011035;%%
  %5 citations counted in INSPIRE as of 01 Jun 2016
  
  %\cite{Verbiest:2016vem}
\bibitem{Verbiest:2016vem} 
  J.~P.~W.~Verbiest {\it et al.},
  %``The International Pulsar Timing Array: First Data Release,''
  Mon.\ Not.\ Roy.\ Astron.\ Soc.\  {\bf 458}, 1267 (2016)
  %doi:10.1093/mnras/stw347
  [arXiv:1602.03640 [astro-ph.IM]].
  %%CITATION = doi:10.1093/mnras/stw347;%%
  %4 citations counted in INSPIRE as of 16 May 2016

    %\cite{Kim:1979if}
\bibitem{Kim:1979if}
  J.~E.~Kim,
  %``Weak Interaction Singlet And Strong CP Invariance,''
  Phys.\ Rev.\ Lett.\  {\bf 43}, 103 (1979).
  %%CITATION = PRLTA,43,103;%%
  
 %\cite{Shifman:1979if}
\bibitem{Shifman:1979if}
  M.~A.~Shifman, A.~I.~Vainshtein and V.~I.~Zakharov,
  %``Can Confinement Ensure Natural CP Invariance Of Strong Interactions?,''
  Nucl.\ Phys.\  B {\bf 166}, 493 (1980).
  %%CITATION = NUPHA,B166,493;%%   

  
  %\cite{Sikivie:1986gq}
\bibitem{Sikivie:1986gq} 
  P.~Sikivie,
  %``The Axion Couplings,''
  UFTP-86-28.
  %%CITATION = UFTP-86-28;%%

  %\cite{Jaeckel:2016qjp}
\bibitem{Jaeckel:2016qjp} 
  J.~Jaeckel, V.~M.~Mehta and L.~T.~Witkowski,
  %``Monodromy Dark Matter,''
  arXiv:1605.01367 [hep-ph].
  %%CITATION = ARXIV:1605.01367;%%

%\cite{Yang:2014}
\bibitem{Yang:2014}
  X.~I.~A.~Yang and R.~Mittal, 
  % ``Acceleration of the Jacobi iterative method by factors exceeding 100 using scheduled relaxation,"
  J.\ Comput.\ Phys.\ {\bf 274}, 695 (2014).
  
  %\cite{Adsuara:2015eds}
\bibitem{Adsuara:2015eds} 
  J.~E.~Adsuara, I.~Cordero-Carri\'on, P.~Cerd\'a-Dur\'an and M.~A.~Aloy,
  %``Scheduled Relaxation Jacobi method: improvements and applications,''
  arXiv:1511.04292 [math.NA].
  %%CITATION = ARXIV:1511.04292;%%



%\cite{Gelmini:1988sf}
\bibitem{Gelmini:1988sf} 
  G.~B.~Gelmini, M.~Gleiser and E.~W.~Kolb,
  %``Cosmology of Biased Discrete Symmetry Breaking,''
  Phys.\ Rev.\ D {\bf 39}, 1558 (1989).
  %doi:10.1103/PhysRevD.39.1558
  %%CITATION = doi:10.1103/PhysRevD.39.1558;%%
  %59 citations counted in INSPIRE as of 09 Jun 2016

%\cite{Gleiser:1998na}
\bibitem{Gleiser:1998na} 
  M.~Gleiser and R.~Roberts,
  %``Gravitational waves from collapsing vacuum domains,''
  Phys.\ Rev.\ Lett.\  {\bf 81}, 5497 (1998)
  %doi:10.1103/PhysRevLett.81.5497
  [astro-ph/9807260].
  %%CITATION = doi:10.1103/PhysRevLett.81.5497;%%
  %19 citations counted in INSPIRE as of 16 May 2016

%\cite{Hiramatsu:2010yz}
\bibitem{Hiramatsu:2010yz} 
  T.~Hiramatsu, M.~Kawasaki and K.~Saikawa,
  %``Gravitational Waves from Collapsing Domain Walls,''
  JCAP {\bf 1005}, 032 (2010)
  %doi:10.1088/1475-7516/2010/05/032
  [arXiv:1002.1555 [astro-ph.CO]].
  %%CITATION = doi:10.1088/1475-7516/2010/05/032;%%
  %21 citations counted in INSPIRE as of 16 May 2016
  
  %\cite{Kawasaki:2011vv}
\bibitem{Kawasaki:2011vv} 
  M.~Kawasaki and K.~Saikawa,
  %``Study of gravitational radiation from cosmic domain walls,''
  JCAP {\bf 1109}, 008 (2011)
  %doi:10.1088/1475-7516/2011/09/008
  [arXiv:1102.5628 [astro-ph.CO]].
  %%CITATION = doi:10.1088/1475-7516/2011/09/008;%%
  %13 citations counted in INSPIRE as of 16 May 2016
  
  %\cite{Hiramatsu:2013qaa}
\bibitem{Hiramatsu:2013qaa} 
  T.~Hiramatsu, M.~Kawasaki and K.~Saikawa,
  %``On the estimation of gravitational wave spectrum from cosmic domain walls,''
  JCAP {\bf 1402}, 031 (2014)
  %doi:10.1088/1475-7516/2014/02/031
  [arXiv:1309.5001 [astro-ph.CO]].
  %%CITATION = doi:10.1088/1475-7516/2014/02/031;%%
  %13 citations counted in INSPIRE as of 16 May 2016
  
    
    %\cite{Janssen:2014dka}
\bibitem{Janssen:2014dka} 
  G.~Janssen {\it et al.},
  %``Gravitational wave astronomy with the SKA,''
  PoS AASKA {\bf 14}, 037 (2015)
  [arXiv:1501.00127 [astro-ph.IM]].
  %%CITATION = ARXIV:1501.00127;%%
  %16 citations counted in INSPIRE as of 16 May 2016


  
\end{thebibliography}
\end{document}